\documentclass[aps,pra,twocolumn,showpacs,superscriptaddress]{revtex4} 
\usepackage{graphicx}
\usepackage{amsmath}
\usepackage{amssymb}
\usepackage{epstopdf}
\usepackage{amsfonts}
\usepackage{dcolumn}
\usepackage{bigints}
\usepackage{xcolor}
\usepackage{hyperref}
\usepackage[normalem]{ulem}
\begin{document}
\title{Finite SSH chains coupled to a two-level emitter: 
Hybridization of edge and emitter states}
\author{C. I. Kvande}
\address{Homer L. Dodge Department of Physics and Astronomy,
  The University of Oklahoma,
  440 W. Brooks Street,
  Norman,
Oklahoma 73019, USA}
\address{Center for Quantum Research and Technology,
  The University of Oklahoma,
  440 W. Brooks Street,
  Norman,
Oklahoma 73019, USA}
\address{Physics Department, Kalamazoo College, 1200 Academy Street, Kalamazoo,
Michigan 49006, USA}
\author{D. B. Hill}
\address{Homer L. Dodge Department of Physics and Astronomy,
  The University of Oklahoma,
  440 W. Brooks Street,
  Norman,
Oklahoma 73019, USA}
\address{Center for Quantum Research and Technology,
  The University of Oklahoma,
  440 W. Brooks Street,
  Norman,
Oklahoma 73019, USA}
\author{D. Blume}
\address{Homer L. Dodge Department of Physics and Astronomy,
  The University of Oklahoma,
  440 W. Brooks Street,
  Norman,
Oklahoma 73019, USA}
\address{Center for Quantum Research and Technology,
  The University of Oklahoma,
  440 W. Brooks Street,
  Norman,
Oklahoma 73019, USA}
\date{\today}

\begin{abstract}
The Hamiltonian for the one-dimensional SSH chain is one of the simplest Hamiltonians that supports topological states. This work considers between one and three finite SSH chains 
with open boundary conditions
that 
either
share a lattice site (or cavity), which---in turn---is coupled to a two-level emitter,
or are
coupled to the same two-level emitter.
We investigate the system properties as functions of the emitter-cavity coupling strength $g$ and the detuning between the emitter energy and the center of the band gap. 
It is found that the energy scale introduced by the edge states that are supported by the uncoupled finite SSH chains leads to a $g$-dependent hybridization of the emitter and edge states 
that is unique to finite-chain systems.
A
highly accurate
analytical three-state model that captures the band gap physics
of $k$-chain ($k \ge 1$) systems
is developed.  
To quantify the robustness of the topological system characteristics, the
inverse participation ratio for the cavity-shared and emitter-shared systems consisting of $k$ chains 
is
analyzed as a  function of the 
onsite
disorder strength.
The $g$-dependent hybridization of the emitter and uncoupled edge states can be probed dynamically.
\end{abstract}
\maketitle

\section{Introduction}
The study of individual photons confined in a reflective cavity interacting with matter, frequently a 
few-level emitter (e.g., an atom), is at the heart of many quantum studies~\cite{carusotto2013,reiserer2015,forndiaz2019,janitz2020,schlawin2022}. Chief accomplishments in the field of cavity quantum electrodynamics (QED), such as the manipulation of atoms through photons and, conversely, the manipulation of individual photons by atoms, were recognized by the 2012 Nobel Prize for Physics~\cite{haroche2013,wineland2013}. An important extension of cavity QED is wave guide QED, where the cavity is replaced by a one-dimensional radiation channel or wave guide~\cite{ciccarello2011,lalumiere2013,zheng2013,calajo2016,bello2019,masson2020,kannon2020,kim2021}. The one-dimensional wave guide 
confines the photons, which  interact with one or more  quantum emitters that are localized at specific positions along the wave guide.
Such systems can feature hybridized bound and propagating light-emitter states as well as super- and sub-radiance and play a central role in various quantum information processing protocols~\cite{petrosyan2008,flamini2019,slussarenko2019,brehm2021}. In dissipation engineering protocols, the wave guide serves as a highly-tunable non-trivial reservoir~\cite{martin2011,mirza2016,kumlin2019}.

\begin{figure}[t]
\vspace*{.2cm}
\hspace*{-0.6cm}
\includegraphics[width=0.65\textwidth]{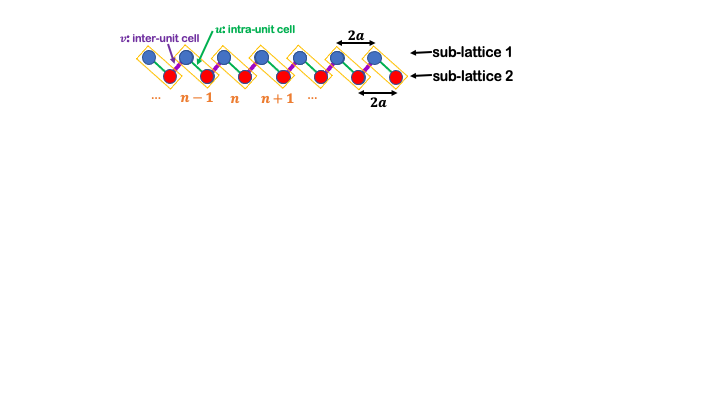}
\vspace*{-5.4cm}
\caption{Illustration of SSH chain.
Each unit cell, marked by an orange box and labeled by $n$, contains
a cavity that belongs to sub-lattice~1 (blue filled 
circle)
and a cavity that belongs to sub-lattice~2 (red filled 
circle).
The intra-unit and inter-unit hopping
energies are denoted by $u$ and $v$, respectively.
The spacing between neighboring unit cells is $2a$.
}
\label{fig_schematic}
\end{figure}

The interplay between a topological wave guide QED bath and one or more localized two-level emitters
(2LEs)
was investigated in a seminal paper by Bello {\em{et al.}} in 2019~\cite{bello2019}. It was found that the coupling of a 
2LE
to a
photonic bath 
with periodic boundary conditions (BCs)
described by the Su-Schrieffer-Heeger (SSH) Hamiltonian gives rise to a chiral 
zero-energy
bound state if the emitter's frequency is tuned to lie in the 
middle of the
bandgap of the bath dispersion. The SSH model was originally introduced to describe solitons in polyacetylene~\cite{su1979} and has been used extensively as an analytically tractable  
model for topological investigations~\cite{capone1997,meier2016,lieu2018,scollon2020}. The SSH bath consists of two sub-lattices (sub-lattice~1 and sub-lattice~2; see Fig.~\ref{fig_schematic}) with inter-unit hopping energy $v$ and intra-unit hopping energy $u$. 
For $|v|>|u|$,
the 
zero-energy
chiral bound state supported by the emitter-chiral wave guide Hamiltonian 
with the emitter tuned to be in resonance with the middle of the band gap,
was found to have the following characteristics
for all emitter-cavity coupling strengths $g$~\cite{bello2019,leonforte2021}:
(i) The photonic component of the bound state has only a finite amplitude in the sub-lattice that the two-level emitter does not couple to (in our case, the emitter couples to sub-lattice 1 of unit cell $n^*$, implying that the photonic component occupies sub-lattice 2).
(ii) The photonic component of the bound state occupies only the side of the chain where the cavity of sub-lattice 2 of the unit cell adjacent to unit cell $n^*$ is connected to the cavity of sub-lattice~1 of the unit cell $n^*$ via a strong bond (left arm in Fig.~\ref{fig_schematic}).
(iii) The bound state 
inherits the properties of the topological edge state, e.g.,
is robust against disorder.

This work considers a {\em{finite}} SSH chain with open BCs coupled to a 
2LE.
Without the 2LE,
the SSH Hamiltonian 
with open BCs
supports two topologically protected edge states that live in the bandgap. Building on the  work presented in the supplemental material of Bello {\em{et al.}}~\cite{bello2019},
the system properties are analyzed as functions of the emitter-bath coupling strength $g$ and the emitter frequency, focusing on parameter combinations for which the 
hybridization
between the emitter and the edge states plays a prominent role.
The finite wave guide with open BCs and zero detuning supports, as the corresponding system with periodic BCs,
a zero-energy state.
The photonic contribution of this state has its maximum at a cavity located at the edges of the chain as opposed to, as found for periodic BCs, at a cavity that sits next to the cavity that the emitter is coupled to.
For $g$ larger than a value that depends on the edge energy of the finite SSH chain and the emitter location on the chain, the emitter contribution to the chiral zero-energy state is essentially zero. The emitter instead hybridizes with the two $g=0$ edge states, resulting in two finite-energy states that play the role of edge states in the arm of the chain that is not occupied by the chiral zero-energy state. A simple analytical three-state (3-state) model that captures the behaviors for both vanishing and non-vanishing detuning is presented.

Motivated by the possibility that cavities can be connected in non-trivial geometries, 
we extend our studies to two and three ``crossed chains" 
that are connected either by sharing a cavity or by coupling to a shared 
2LE.
Both scenarios can be thought of
as having a single site, either the shared cavity or the shared emitter, with a coordination number that is, 
respectively,
two and three times larger for the 
2-chain and 3-chains
scenarios
than the coordination number of the other cavities.
Even though the number of  $g=0$ states in the middle of the gap increases with $k$, we find that the $g$-dependent characteristics of emitter- and cavity-shared $k$-chain systems can be described by the same analytical 3-state Hamiltonian as the 1-chain system, provided the effective coupling constant is chosen accordingly. Our results extend readily to $k>3$.

Our findings highlight that finite baths, as frequently realized experimentally, display---compared to infinite baths---distinct characteristics that are due to the new energy and length scales introduced by the finiteness of the system. In the case of the topological bath, there is not only an energy scale that emerges from the finite length of the chain (which sets, e.g., a time for photons to travel to the end of the chain and back
and which also exists for non-topological chains) but 
also 
an 
energy scale that emerges from the splitting of the two edge states supported by the bath 
(this energy scale does not exist in a non-topological bath and goes to zero for a topological bath as $N$ approaches infinity).

The remainder of this paper is organized as follows.
Section~\ref{sec_theory} introduces the system Hamiltonians and static properties, including a discussion of the eigen spectra of $k$-chain
 systems ($k=1-3$) and the emergence of dark states; mathematical details are relegated to Appendices~\ref{sec_appendix1}-\ref{sec_appendix4}. 
Section~\ref{sec_ipr} investigates the robustness of the states that have topological characteristics to chiral-symmetry-breaking disorder while Sec.~\ref{sec_dynamics} shows that the $g$-dependent hybridization can be probed dynamically.
Section~\ref{sec_summary}
concludes and presents an outlook.

\section{Hamiltonian,
eigen energies, and eigen states}
\label{sec_theory}

\subsection{SSH Hamiltonian}
\label{sec_theory_part1}
For a chain that consists of $N$ unit cells,
the SSH Hamiltonian
$\hat{H}_{\text{SSH}}$, which is schematically illustrated in Fig.~\ref{fig_schematic},  reads
\begin{eqnarray}
\label{eq_ham_ssh}
\hat{H}_{\text{SSH}} = 
\sum_{n=1}^{N-1}
\left[
\left(
u \hat{c}_{n,1}^{\dagger} \hat{c}_{n,2}
+
v \hat{c}_{n,2}^{\dagger} \hat{c}_{n+1,1}
\right)
+ h.c.
\right] + \nonumber \\
\left[ \left(
u \hat{c}_{N,1}^{\dagger} \hat{c}_{N,2}
+
v_N \hat{c}_{N,2}^{\dagger} \hat{c}_{1,1}
\right)
+ h.c. \right]
,
\end{eqnarray}
where $\hat{c}_{n,j}$
annihilates an excitation in the $j$th sub-lattice ($j=1$ or $2$) of the $n$th unit cell.
The parameters $u$ and $v$, which are taken to be real and positive throughout this paper, denote intra-unit cell and inter-unit cell hopping energies, respectively.
Periodic and open BCs are realized for
$v_N=v$ and $v_N=0$, respectively.
Throughout, we  
use 
$u$ and $\hbar/u$ as our
energy and  time units.
The SSH model has elucidated phenomena in many sub-disciplines of physics including chemical physics~\cite{perebeinos2005}, condensed matter physics~\cite{stjain2017}, cold atom physics~\cite{meier2016}, and relativistic field theories~\cite{jackiw1981}.
Throughout this paper, we have a scenario in mind where each unit cell contains two  cavities (one that belongs to sub-lattice~1 and one that belongs to sub-lattice~2; see Fig.~\ref{fig_schematic}) and where the  operators 
$\hat{c}_{n,j}$ and $\hat{c}_{n,j}^{\dagger}$
annihilate and create a photon in the cavity that belongs to sub-lattice~$j$ of the $n$th unit cell. 

Since the SSH Hamiltonian
possesses a chiral symmetry, it is a paradigmatic model for studying topology.
Specifically, the chiral operator
$\hat{C}$ and $\hat{H}_{\text{SSH}}$ anti-commute,
$\hat{C} \hat{H}_{\text{SSH}} \hat{C} = -
\hat{H}_{\text{SSH}}$,
where $\hat{C}$ is defined in terms of the
projection operators
$\hat{P}_j$ ($j=1$ or $2$),
$\hat{C}=\hat{P}_1-\hat{P}_2$
and
$\hat{P}_j = 
\sum_{n=1}^N 
\hat{c}_{n,j}^{\dagger} \hat{c}_{n,j}$.
For concreteness, we consider the set-up in Fig.~\ref{fig_schematic}. If $v$ is smaller than $u$,
$\hat{H}_{\text{SSH}}$ is topologically trivial. 
Since the inter-unit hopping strength is weaker than the intra-unit hopping strength, 
the two cavities contained in a given unit cell are ``binding together", i.e., the hopping strengths ``respect" the chain's division into unit cells.
If, on the other hand,
$v$ is larger than $u$, $\hat{H}_{\text{SSH}}$ is topologically non-trivial.
In this case, a cavity from the $n$th unit cell and a cavity from the 
$(n+1)$th unit cell are binding together, leading to inter-unit cell bonds.
For open BCs, this leads to a single dangling or unpaired cavity
on each end of the chain and the emergence of two edge states
that are predominantly located at the first and $N$th unit cells~\cite{asboth}.

For later reference, we introduce approximate expressions for the  edge states $| \psi_{\pm}^{\text{C}1} \rangle$:
 \begin{eqnarray}
 \label{eq_b1}
 |\psi_{\pm}^{\text{C}1} \rangle
 = \frac{1}{\sqrt{2}}
 \left(
 | \psi_{\text{edge,L}}^{\text{C}1} \rangle \pm | \psi_{\text{edge,R}}^{\text{C}1} \rangle
 \right),
 \end{eqnarray}
 where $|\psi_{\text{edge,L}}^{\text{C}1}\rangle$ and $|\psi_{\text{edge,R}}^{\text{C}1}\rangle$
 are localized in sub-lattice~1 near the first unit cell and in sub-lattice~2 near the $N$th unit cell, respectively:
 \begin{eqnarray}
 \label{eq_psi_edgeL_ring1}
 |\psi_{\text{edge,L}}^{\text{C}1} \rangle=
  \sum_{n=1}^N
 c_{n,1} |n,1 \rangle
 \end{eqnarray}
 and
 \begin{eqnarray}
 \label{eq_psi_edgeR_ring1}
 | \psi_{\text{edge,R}}^{\text{C}1} \rangle=
  \sum_{n=1}^N
 c_{n,2} |n,2 \rangle.
 \end{eqnarray}
 Here, the site basis states $|n,j\rangle$, where $n=1,\cdots,N$ labels the unit cell and $j=1,2$ indicates the sub-lattice, are used.
 The superscript ``C$1$" (chain 1) is introduced with view toward the $k>1$ discussion  below.
 The expansion coefficients $c_{n,1}$ and $c_{n,2}$  read
 \begin{eqnarray}
 c_{n,1} = {\cal{N}} (-1)^{n+1}   \left( u/v \right)^{n-1}
 \end{eqnarray}
 and
 \begin{eqnarray}
 c_{n,2} = {\cal{N}} (-1)^{N-n}   \left(u/v\right)^{N-n},
 \end{eqnarray}
 with ${\cal{N}}$ denoting a normalization constant, 
 \begin{eqnarray}
 \label{eq_b7}
 {\cal{N}}=
 \left[1-
 \left(u/v\right)^2 
 \right]^{1/2}  \left[ 1-
 \left( u/v \right)^{2N}
 \right]^{-1/2}
 .
 \end{eqnarray}
 Equations~(\ref{eq_b1})-(\ref{eq_b7}) become exact in the $N \rightarrow \infty$ limit.
The states $| \psi_{\pm}^{\text{C}1} \rangle$ have energy
$\pm E_{\text{edge}}$, where 
 \begin{eqnarray}
 \label{eq_b8}
 E_{\text{edge}}= (-1)^{N+1}
 {\cal{N}}^2 
\left( u/v \right)^{N-1} 
 u.
 \end{eqnarray}
The energy $E_{\text{edge}}$ approaches zero exponentially with increasing $N$.

In the $N \rightarrow \infty$ limit, the edge states have vanishing energy
and are characterized by a localization length $\zeta_{\text{loc}}=2 a/\ln(v/u)$, where $2a$ denotes the 
separation between neighboring unit cells~\cite{asboth}.
For $v=2u$, as considered throughout this paper, $\zeta_{\text{loc}}$ evaluates to $\approx 2.89a$.
 Moreover, if a site from sub-lattice~1
(sub-lattice~2) sits at the end of the chain, the edge state has zero amplitude in sub-lattice~2 (sub-lattice~1) at that end. 
Since
  zero-energy eigen states are simultaneously eigen states of $\hat{C}$, this
 follows directly from the chiral symmetry.
 Note that zero-energy eigen states, and correspondingly edge states, are not supported for periodic BCs. Yet, the systems with open and periodic BCs are intimately related through the bulk-edge correspondence~\cite{asboth,hasan2010,chiu2016}. 
 
\begin{figure}[t]
\includegraphics[width=0.47\textwidth]{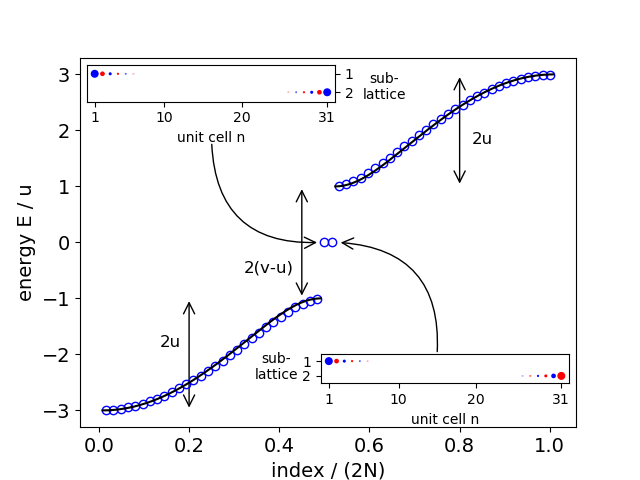}
\caption{Characteristics of single finite SSH 
chain
(without the emitter) with open BCs, $N=31$, and $v/u=2$.
The 
blue circles in the
main panel 
show
the eigen energies as a function of the normalized eigen state index. 
The eigen energies are distributed symmetrically around zero energy. The spectrum features two nearly continuous energy bands and two states in the band gap with energy close to zero. 
The energy gap has an energy width of $2(v-u)$ while the bands each have a width of $2u$; correspondingly, the lowest and highest energy levels are separated by $2(u+v)$.
For comparison, the black solid lines show the energy bands of the infinite chain with periodic BCs; the blue circles agree quite well with the solid lines.
The eigen states of the two edge states 
for finite $N$
are illustrated in the upper left  and the lower right insets. The size of the filled circles is directly proportional to the square of the expansion coefficient of  the site basis state $|n,j\rangle$, with the color marking the sign of the expansion coefficients
(see text for details). }
\label{fig_energy_ssh}
\end{figure}

The blue circles in Fig.~\ref{fig_energy_ssh} show
  the energy spectrum, plotted as a function of the normalized eigen state index,
 for a chain with open BCs, $v=2u$, and $N=31$.
 In this case, the energy
 $\pm E_{\text{edge}}$
 of the states in the band gap is $\pm 6.98\times 10^{-10} u$;
 within the digits reported, this agrees with the approximate expression (\ref{eq_b8}). For comparison, the black solid  line shows the eigen energies for the infinite chain with periodic BCs. The agreement between the finite $N$ and infinite $N$ energy bands is very good.
 The 
 numerically determined
 edge states are illustrated in the insets in the upper left and lower right of Fig.~\ref{fig_energy_ssh} using the 
 site basis states
 $|n,j\rangle$.
  The size of the filled circles in Fig.~\ref{fig_energy_ssh}
 is directly proportional to the square of the amplitude of the expansion coefficients in  sub-lattice~1 (upper row) and sub-lattice~2 (lower row) in the $n$th unit cell; the color of the circles (blue and red) represents the sign (positive and negative) of the expansion coefficients.
Despite
the finite number of unit cells, the 
eigen states inherit the key characteristics
 of the thermodynamic system ($N \rightarrow \infty$ 
 limit), i.e., the edge states have finite amplitude on just one sub-lattice. Moreover, the localization length  of the finite-chain edge states is very close to the localization length $\zeta_{\text{loc}}$ for infinite $N$.

\subsection{
Single
SSH chain coupled to emitter
}
\label{sec_theory_part2}

\begin{figure}[t]
\includegraphics[width=0.65\textwidth]{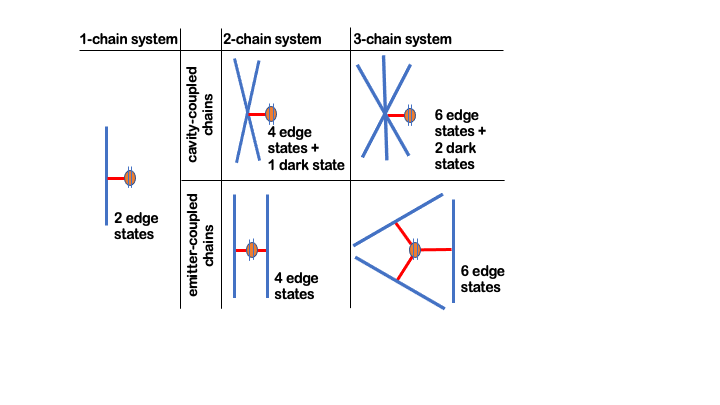}
\vspace*{-2.cm}
\caption{Schematic of emitter- and cavity-shared 
$k$-chain
systems.
The blue lines represent SSH chains; from the 
left to the right,
the number of chains increases from one to three. Each chain is coupled to an emitter (orange circle with two 
vertical
lines). The red line represents the emitter-cavity coupling. In the emitter-shared scenario 
(top middle and top right panels), 
chains
do not share cavities and one cavity of each 
chain
is coupled to the emitter; for $g=0$, the chains in the 
2- and 3-chain
cases are decoupled.
In the cavity-shared scenario 
(bottom middle and bottom right panels),
the 
2- and 3-chain 
systems share the cavity that the emitter is coupled to.
For the 1-chain system, the number of sites is $2N$ (there exists no distinction between the emitter- and cavity-coupled 1-chain systems).
The number of sites is
$4N$ and $6N$ for the emitter-shared 
$2$- and $3$-chain
systems, respectively.
The number of sites is
$4N-1$ and $6N-2$ for the cavity-shared 
$2$- and $3$-chain
systems, respectively.
The number of edge and dark states reported next to the schematic applies to the cavity part of the Hilbert space, excluding the emitter; in the special case where the emitter detuning is zero, the coupled systems support one additional dark state (see Appendix~\ref{sec_appendix1}).
}
\label{fig_schematic_ring}
\end{figure}

As alluded to in the introduction, we are interested in 
$k$-chain
systems ($k=1-3$) coupled to a single 
2LE
with ground state $|g \rangle$ 
(energy $0$) and excited state $|e\rangle$ (energy $\hbar \omega_e$). 
Schematics of these systems are shown in Fig.~\ref{fig_schematic_ring}. This section introduces the 
1-chain
Hamiltonian $\hat{H}_{\text{C1-2LE}}$, 
which is written as a sum of the SSH chain, the 
2LE
Hamiltonian 
$\hat{H}_{\text{2LE}}$,
and the coupling term
$\hat{H}_{\text{int}}$ (see, e.g., Ref.~\cite{ciccarello2011}),
\begin{eqnarray}
\label{eq_ham_1ring}
\hat{H}_{\text{C1-2LE}}=
\hat{H}_{\text{SSH}}+
\hat{H}_{\text{2LE}}+
\hat{H}_{\text{int}},
\end{eqnarray}
\begin{eqnarray}
\label{eq_ham_emitter}
\hat{H}_{\text{2LE}}=
\frac{\hbar \omega_e}{2}
\left(\hat{\sigma}^z
+1 \right),
\end{eqnarray}
and
\begin{eqnarray}
\label{eq_ham_int}
\hat{H}_{\text{int}}=
g \left( \hat{c}_{n^*,1}^{\dagger} \hat{\sigma}^-
+
\hat{c}_{n^*,1} \hat{\sigma}^+
\right).
\end{eqnarray}
The operators
$\hat{\sigma}^z$,
$\hat{\sigma}^+$,
and
$\hat{\sigma}^-$, which act in the Hilbert space of the 
2LE,
read
$\hat{\sigma}^z=|e\rangle \langle e|-|g\rangle \langle g|$,
$\hat{\sigma}^+=|e\rangle \langle g|$,
and
$\hat{\sigma}^-=|g\rangle \langle e|$.
In Eq.~(\ref{eq_ham_int}),
the emitter couples,
with coupling energy $g$ ($g \ge 0$), to the 
cavity 
of sub-lattice~1 that belongs to the $(n^*)$th unit cell.
In the examples considered in this work,
$N$ is odd and $n^*$ is equal to $(N+1)/2$.
Coupling to a cavity that belongs to sub-lattice~2 can be treated in the same way and yields analogous results.
Since Eq.~(\ref{eq_ham_int}) employs the rotating wave approximation~\cite{ref_petro-book}, we restrict ourselves to $g/u \le 0.1$ throughout this work.

The Hamiltonian $\hat{H}_{\text{C1-2LE}}$
commutes with the excitation operator
$\hat{N}_{\text{exc}}$,
$\hat{N}_{\text{exc}}=
\hat{P}_1 + \hat{P}_2 + |e \rangle \langle e |$,
and can thus be diagonalized separately for each 
excitation manifold~\cite{bello2019}. 
Since the dynamics discussed in 
Sec.~\ref{sec_dynamics}
start with the emitter in state $|e \rangle$ and the SSH chain in the zero-photon vacuum state $|\text{vac} \rangle$ (this state has $\langle \hat{N}_{\text{exc}} \rangle=1$),
we are interested in the single-excitation manifold, which is spanned by the basis states
$|n,j;g \rangle$, 
where the first two entries refer to the SSH chain ($n=1,\cdots,N$ and $j=1,2$) and the last entry refers to the emitter, and $| \text{vac};e\rangle$. 
Since $\hat{C} \hat{H}_{\text{C1-2LE}} \hat{C}$ is not
equal to 
$-\hat{H}_{\text{C1-2LE}}$, the introduction of the emitter leads to a breaking of the chiral symmetry: 
the emitter can be thought of as a 
chiral symmetry-breaking perturbation.

\begin{widetext}

\begin{figure}[t]
\includegraphics[width=0.85\textwidth]{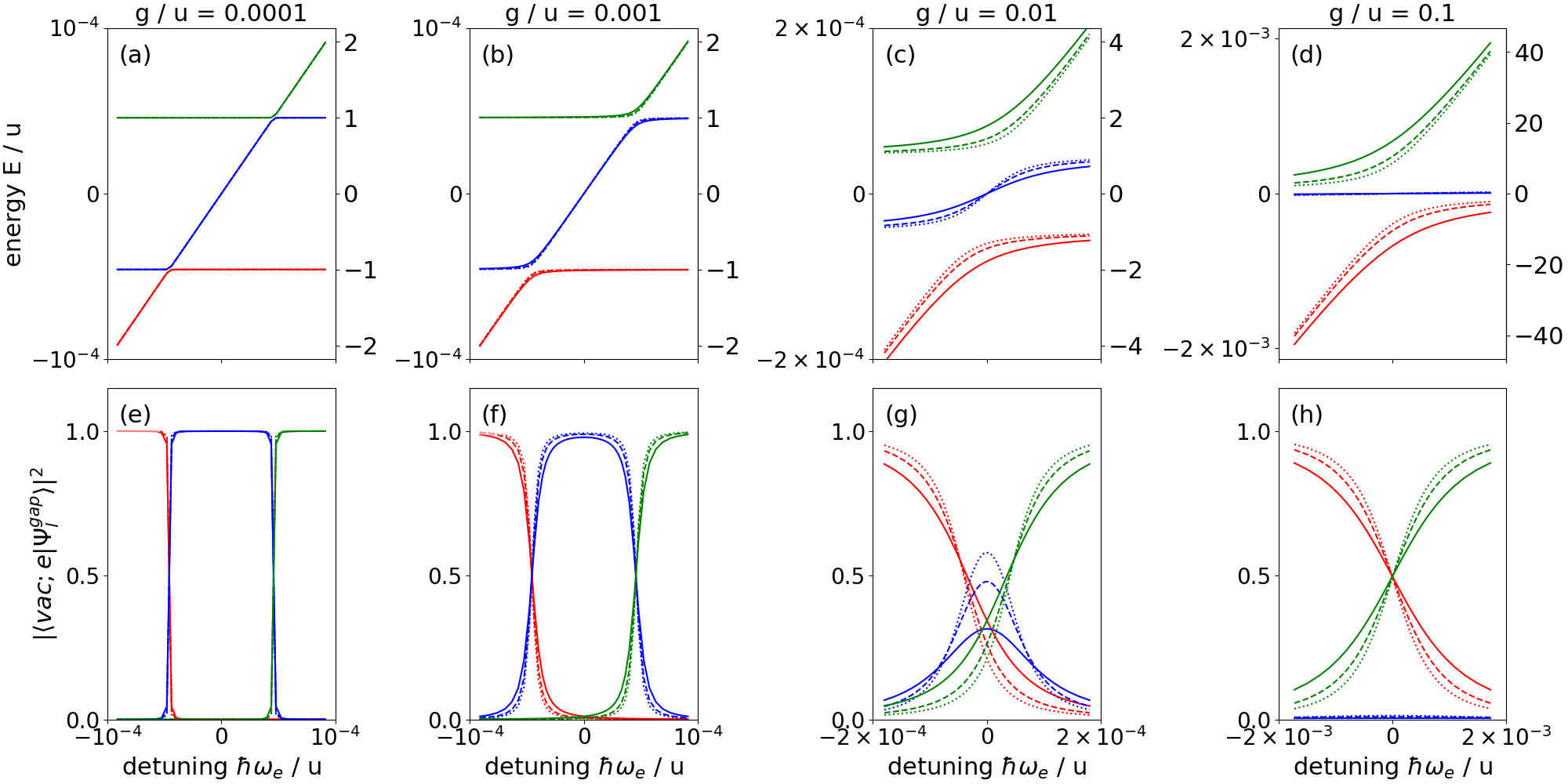}
\caption{Static properties of the 
1-chain, cavity-shared 2-chain, and cavity-shared 3-chain
systems with open BCs, $N=15$, $n^*=8$, and $v/u=2$ as a function of the detuning $\hbar \omega_e$; the figure zooms in on the physics in the vicinity of the middle of the gap. 
The solid, dashed, and dotted lines are for the 
1-chain, 2-chain, and 3-chain
systems, respectively. The coupling $g/u$ increases from left to right; specifically,
the first, second, third, and fourth columns are for 
$g/u= 10^{-4}$,
$10^{-3}$,
$10^{-2}$
and
$10^{-1}$, respectively.
The top row shows the three energy levels $E_l^{\text{gap}}$ located in the energy gap.
The left axis reports the energy in units of $u$ ($u$ is much larger than the typical scale of the energies in the gap) while the right axis reports the energy in units of $E_{\text{edge}}$ ($E_{\text{edge}}\approx 4.58 \times 10^{-5}u$).
The bottom row shows the
overlap square ${\cal{O}}_l$.
Note that the scale of the figures is the same for the first and second columns but changes for the third and fourth columns.
}
\label{fig_energy_ring}
\end{figure}  

\end{widetext}

Solid lines in the top row of Fig.~\ref{fig_energy_ring} show the near-zero eigen energies of $\hat{H}_{\text{C1-2LE}}$
 as a function of the emitter energy $\hbar \omega_e$ for $v/u=2$, $N=15$, $n^*=8$, and four different  $g/u$, namely $g/u=10^{-4}-10^{-1}$. The spectrum is calculated using open BCs.
The emitter energy can be interpreted as a detuning from the center of the band gap.
For $g=0$, the eigen states of the three eigen energies shown in Fig.~\ref{fig_energy_ring}
correspond to the two edge states 
$|\psi_-^{\text{C}1};g\rangle$
and 
$|\psi_+^{\text{C}1};g\rangle$ 
and the excited emitter
state $|\text{vac};e\rangle$.
As the coupling $g$ is turned on, these three states mix and the corresponding eigen energies undergo avoided crossings. 
The eigen energies that belong to the two nearly continuous bands (not shown) and their eigen states, in contrast, remain essentially unchanged.
For small $g/u$, 
avoided crossings 
between two states occur when the detuning (or emitter energy) $\hbar \omega_e$ is equal to the energy $\pm E_{\text{edge}}$ of the edge states supported by $\hat{H}_{\text{SSH}}$ (for the $N=15$ system considered in Fig.~\ref{fig_energy_ring}, $E_{\text{edge}}\approx 4.58 \times 10^{-5}u$). As expected, 
both
avoided crossings 
become broader with increasing $g/u$.
The two avoided crossings start to overlap (implying hybridization of three states)
for $g/u \gtrsim 10^{-2}$.
For $g/u=0.1$, Fig.~\ref{fig_energy_ring}(d)  suggests that the green and red energy levels undergo an avoided crossing, with the energy level shown in blue being 
decoupled and having, on the scale shown, zero energy.

Denoting the three eigen states whose energies lie in the gap by $|\psi_l^{\text{gap}}\rangle$ ($l=1-3$), we find that the initial state
$|\text{vac};e\rangle$ considered in the dynamical studies discussed in 
Sec.~\ref{sec_dynamics}
can be decomposed with good accuracy as
\begin{eqnarray}
\label{eq_initialstate_3state}
|\text{vac};e\rangle
\approx
\sum_{l=1}^3
d_l^{\text{gap}} |\psi_l^{\text{gap}} \rangle,
\end{eqnarray}
where 
$d_l^{\text{gap}}=
\langle   \psi_l^{\text{gap}} |
\text{vac};e \rangle$.
Solid lines in the second row of Fig.~\ref{fig_energy_ring} show the overlap
square $O_l$, 
$O_l=|\langle \text{vac};e | \psi_l^{\text{gap}} \rangle|^2$.
Since the quantity $\sum_{l=1}^3 {\cal{O}}_l$ is greater than 
$0.996$
for all 1-chain systems considered in Fig.~\ref{fig_energy_ring}, 
the results presented in the second row of Fig.~\ref{fig_energy_ring} allow us to forecast where population transfer is expected since population transfer occurs only if the initial state projects onto two or more eigen states of the coupled system.
%

Since the three gap states $|\psi_l^{\text{gap}}\rangle$ can be written, with good accuracy, as a superposition of
the uncoupled approximate $g=0$ states 
$|\psi_-^{\text{C}1};g\rangle$,
$|\psi_+^{\text{C}1};g\rangle$,
and $|\text{vac};e\rangle$ for all parameter combinations considered in Fig.~\ref{fig_energy_ring}, we use them
 to construct
 the 3-state Hamiltonian matrix
 $\underline{H}_{\text{3-st.}}(G)$: 
 \begin{eqnarray}
 \label{eq_1ring_matrix}
 \underline{H}_{\text{3-st.}}(G)=
 \left(
 \begin{array}{ccc}
 -E_{\text{edge}} & 0 & G \\
 0 & E_{\text{edge}} & G \\
 G & G & \hbar \omega_e \\
 \end{array}
 \right),
 \end{eqnarray}
 where the effective coupling energy $G$ is defined through  $G=\langle \psi_{\pm}^{\text{C}1};g|\hat{H}_{\text{int}}|\text{vac};e\rangle$; note, $G$ is real. 
 Using the analytical expressions given in Eqs.~(\ref{eq_b1})-(\ref{eq_b7}), we find
 \begin{eqnarray}
 \label{eq_b10}
 G = 
  g c_{n^*,1} / \sqrt{2}.
 \end{eqnarray}
 While Eq.~(\ref{eq_1ring_matrix}) is characterized by the effective coupling constant $G$, the 3-state model introduced in the supplemental material of Ref.~\cite{bello2019} contains both $G$ and $-G$. 
 Our 3-state model 
 reproduces the 
1-chain
 gap energies and overlap square data shown 
 in Fig.~\ref{fig_energy_ring} to 
 $0.5$~\% 
 or better.
 Correspondingly, the model serves as
 a highly reliable  tool for understanding the gap physics for the parameter regime of interest in this paper. 
 Since $\hat{H}_{\text{3-st.}}(G)$ lives
  in the space that is spanned by states that have non-vanishing amplitude on sub-lattice~1 only in the left arm of the SSH chain and non-vanishing amplitude on sub-lattice~2 only in the right arm of the SSH chain and since it describes the gap states $|\psi_l^{\text{gap}}\rangle$ accurately,
 the gap states inherit
 chiral characteristics for all parameter combinations shown in Fig.~\ref{fig_energy_ring}.

The 3-state model predicts that the hybridization of the three uncoupled basis states occurs at $|G/  E_{\text{edge}}| \approx 1$ (see Appendix~\ref{sec_appendix2}).
 For the parameters of Fig.~\ref{fig_energy_ring}, this corresponds to $g/u\approx 10^{-2}$. For fixed $u/v$, the transition moves to smaller $g/u$ with increasing $N$ [and, as before, $n^*=(N+1)/2$].
 For fixed $N$ and $n^*=(N+1)/2$, the transition moves to larger $g/u$ with decreasing $v/u$ (keeping $v>u)$.

\begin{figure}[t]
\includegraphics[width=0.47\textwidth]{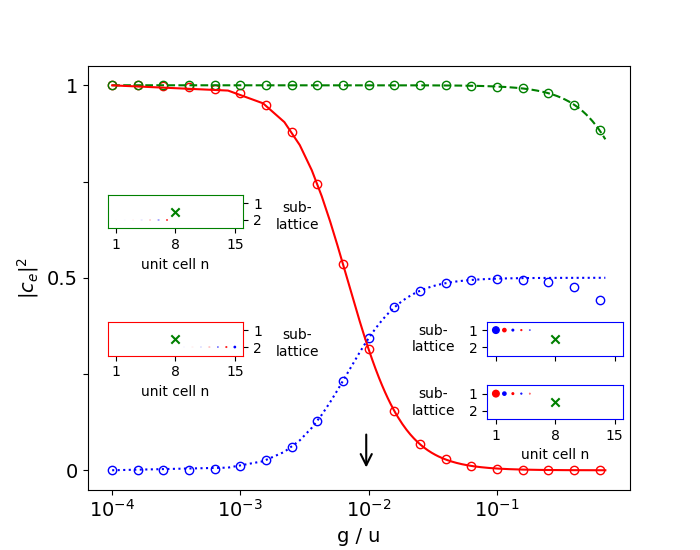}
\caption{
Contribution $|c_e|^2$ of the state $|\text{vac};e\rangle$ to the gap states as a function of $g/u$ for the 1-chain system with $N=15$, $n^*=8$, $v/u=2$, and $\hbar \omega_e=0$; open and periodic BCs are considered.
The lines show approximate analytical results while the symbols show results obtained by diagonalizing the full Hamiltonian. The red solid line and symbols show $|c_e|^2$ of the zero-energy state for open BCs. 
The inset in the lower left (red box) represents the corresponding eigen state for $g / u = 10^{-4}$.
For comparison, the green dashed line and symbols show $|c_e|^2$ for the zero-energy state for periodic BCs.
The inset in the upper left (green box) represents the corresponding eigen state for $g / u = 10^{-4}$.
The blue dotted line and symbols show $|c_e|^2$ of the finite-energy states for open BCs. 
The insets in the lower right corner (blue boxes) represent the corresponding eigen states for $g / u = 10^{-1}$. Note that the sketches of the photonic populations for $g/u=10^{-4}$
are, to enhance readibility, multiplied by $10^6$ (upper green box on the left) and $10^2$ (lower red box on the left) relative to those for $g/u=10^{-1}$. The green cross in the insets marks the unit cell that the emitter is coupled to.
 As a reference point,
the arrow marks the coupling strength for which $|G/E_{\text{edge}}|=1$.
}
\label{fig_new}
\end{figure}

To compare the 1-chain systems with open and periodic BCs for $v/u=2$, $N=15$, and $n^*=8$ (same parameters as used in Fig.~\ref{fig_energy_ring}), we analyze the zero-energy state, which exists for $\hbar \omega_e=0$ for both open and periodic BCs. Red solid and green dashed lines  in Fig.~\ref{fig_new} show the
probability $|c_e|^2$ of the zero-energy state to be in state $|\text{vac};e\rangle$
(approximate analytical expressions) as a function of $g/u$ for open and periodic boundary conditions, respectively.
For comparison, the symbols are obtained by diagonalizing the Hamiltonian $\hat{H}_{\text{C1-2LE}}$.
The agreement between the lines and symbols is excellent. 
For open BCs,
$|c_e|^2$ is close to $1$ for small $g/u$ and drops to a value close to $0$ around $g/u$ values for which $|G/E_{\text{edge}}| \approx 1$ (arrow in Fig.~\ref{fig_new}).
For periodic BCs, in contrast, $|c_e|^2$ 
  does not decrease notably till $g/u$ takes  values of order one (or, more generally, when the coupling energy $g$ becomes comparable to the width of the $g=0$ energy gap). The fact that the coupling strength where the contribution $|c_e|^2$ to the zero-energy state drops significantly differs for open and periodic BCs explicitly demonstrates the role played by $E_{\text{edge}}$. 
 We note that the behaviors of systems with $g/u$ values larger than $0.1-0.5$ need to be interpreted with the understanding that beyond-the-rotating-wave-approximation terms may play a non-negligible role.
 For small $g/u$, the photonic contribution to the zero-energy state of the systems with open and periodic BCs is located on different arms and localized at different positions, namely, as far away from the emitter as possible for open BCs (see the inset in upper left corner) and on cavities close to the emitter for periodic BCs (see the inset in the middle left).
 The blue dotted lines and circles show $|c_e|^2$ for the non-zero energy states of the finite-chain system: $|c_e|^2$
 increases relatively sharply at $|G/E_{\text{edge}}| \approx 1$. For $g/u\gtrsim 0.2$, the agreement between the numerical results and the approximate analytical expression deteriorates. This is not surprising since this is the regime where $g$ is strong enough to couple to states that are not part of $\hat{H}_{\text{3-st.}}$. The insets in the lower right corner show the corresponding eigen states; both have finite photonic contributions on the left arm of the chain.
For finite detuning, direct comparisons between the systems with open and periodic BCs are less straightforward since the energy of the gap states (system with open BCs) and bound state (system with periodic BCs) changes differently with finite detuning.

\subsection{2- and 3-chain systems}
\label{sec_theory_part3}

The cavity- and emitter-shared 
$2$- and $3$-chain
systems are illustrated in Fig.~\ref{fig_schematic_ring}. To make connections between the physics of the 
2- and 3-chain
systems and that of the 
single SSH chain system discussed in the previous section, we start with $g=0$ and then consider what happens for finite $g$ values. 

The emitter-shared 
2- and 3-chain
systems
reduce, for $g=0$, to two and three independent copies of the single SSH 
chain.
 Correspondingly, the 
 2- and 3-chain
 systems with open BCs support a total of four and six edge states, respectively. The energy degeneracy of the edge states is two (three) for the 
 2-chain (3-chain)
 systems:
For the 
2-chain (3-chain)
systems,
two (three) states have energy $E_{\text{edge}}$
and
two (three) states have energy $-E_{\text{edge}}$.
Forming appropriate linear combinations of the degenerate states (see Appendix~\ref{sec_appendix3} for details), we find that four (six) of these states are, to a very good approximation, not affected by the coupling between the cavities and the emitter, i.e., their energies for finite $g$ are approximately equal to $\pm E_{\text{edge}}$. 
The other three energies near zero 
with eigen states  $|\psi_{l}^{\text{gap}} \rangle$
are, as in the case of the 
1-chain
system, very well described by a 3-state model. 
Specifically,
 the 3-state model discussed in
the previous section applies also to the emitter-shared 
$k$-chain
($k>1$) system, provided $g$ is not too large and provided the effective coupling constant is replaced by $\sqrt{k}G$  
(see Appendix~\ref{sec_appendix3} for details). 
We note also that the
initial state  $|\text{vac};e\rangle$ can, to a good approximation, be expanded in terms of 
the three gap states $|\psi_l^{\text{gap}}\rangle$.
The quantity $\sum_{l=1}^3 {\cal{O}}_l$ is greater than $0.993$
for the emitter-shared 2- and 3-chain systems for the parameter combinations covered in Fig.~\ref{fig_energy_ring} (note, though, that the figure,  is for the cavity-shared systems).

\begin{figure}[t]
\includegraphics[width=0.47\textwidth]{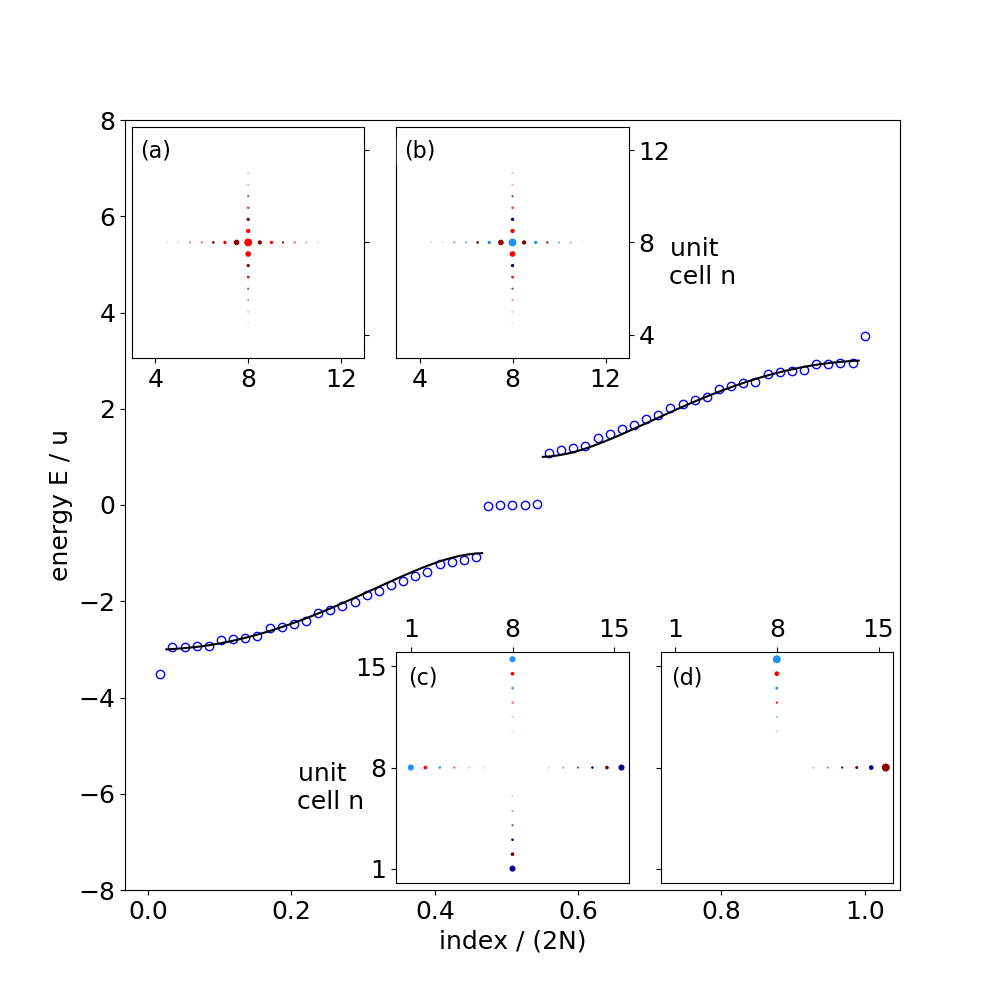}
\caption{Characteristics of the cavity-shared 
2-chain
system (without the emitter) with open BCs, $N=15$, and $v/u=2$; the shared cavity belongs to sub-lattice~1 and is part the $8$th unit cell 
(i.e., $n^*=8$).
The main panel shows the eigen energies as a function of the normalized eigen state index. 
The energy spectrum features two nearly continuous energy bands, five states in the band gap with energy equal to or close to zero  (the state with energy exactly equal to zero is a dark state), one non-topological bound state below the bottom of the lower energy band, and one non-topological bound state above the top of the upper energy band. 
Insets: (a)/(b) The eigen states of the two non-topological bound states are shown in the upper left and upper right insets; they have energy $\pm 3.516 u$. 
(c) 
The edge state 
shown in the lower left inset
has an energy of $-4.56 \times 10^{-5} u$. 
(d) The dark state
shown in  the lower right inset
has an energy of zero.
 The first and second SSH chains are  chosen to lie along the $x$- and $y$-axes (the axes are chosen arbitrarily).
As in Fig.~\protect\ref{fig_energy_ssh}, positive and negative expansion coefficients are shown by blue and red circles, respectively, with the size of the circles being proportional to the square of the expansion coefficients.
As opposed to off-setting
the sites that belong to sub-lattices~1 and 2
(as in Fig.~\ref{fig_energy_ssh}),
the lighter blue and lighter red colors correspond to sites that belong to sub-lattice~1 while the darker blue and darker red colors correspond to sites that belong to sub-lattice~2. Note the different scales of the axes of the insets. 
}
\label{fig_energy_2ring}
\end{figure}

 The cavity-shared 
 2- and 3-chain
 systems are, even for $g=0$, distinct from the 
 1-chain
 system.
 Figure~\ref{fig_energy_2ring} shows the energy of the 
 2-chain
 system with open BCs as a function of the normalized state index for  $N=15$ and $v/u=2$. Since the two chains share one cavity, the total number of sites of the 
 2-chain
 system is $4N-1$ (recall, $N$ refers to the number of unit cells of one of the SSH chains). 
As expected, the energy spectrum consists of two nearly continuous energy bands that are separated by an energy gap, which  supports states with energy close to zero. The number of states in the gap is not four, as might be expected naively by doubling the number of edge states supported by the 
1-chain
system, but five. Four states have finite energy and one state has vanishing energy. 
The latter 
is a delocalized dark state (see Appendix~\ref{sec_appendix1}), which has non-vanishing amplitude in both sub-lattices (see the lower right inset of Fig.~\ref{fig_energy_2ring}). An analogous  non-topological dark state also exists for the 
2-chain
system with periodic BCs but does not exist for the emitter-shared 
2-chain
systems 
with periodic and open BCs.
The other four states with energy close to zero can be divided into two pairs. The states belonging to the first pair, with energy 
$\pm \epsilon$,
are approximately unaffected when $g$ is turned on.
The states belonging to the second pair with energy $\pm E_{\text{edge}}$ (see the lower left inset of Fig.~\ref{fig_energy_2ring} for an example), couple to the emitter and form, together with the state $| \text{vac};e \rangle$, the basis for a 3-state model
(same 
3-state Hamiltonian matrix
as discussed above for the 
1-chain
system, but with $G$ replaced by $G/\sqrt{2}$;
see Appendix~\ref{sec_appendix4}).

For comparison, the cavity-shared 
3-chain
system with open BCs ($6N-2$ sites with one cavity shared by all three 
chains)
supports two dark states 
for $g=0$
whose energy is exactly zero as well as another six states that also reside 
in the energy gap. The existence of these six states might be expected based on the naive 
argument that the number of edge states supported by the 
1-chain
system triples for the 
3-chain
system.
The six states can be divided into two groups, four states that have energy $\pm \epsilon$ (two states with positive energy and two states with negative energy) and two states that have energy $\pm E_{\text{edge}}$. The latter two states couple to the emitter when $g$ is finite
 and are described well by the 3-state 
Hamiltonian matrix $\underline{H}_{\text{3-st.}}(G/\sqrt{3})$ (see Appendix~\ref{sec_appendix4}).

The energy spectrum  for the cavity-shared 
2-chain
system shown in Fig.~\ref{fig_energy_2ring}
features one  energy state below the bottom of the lower band and one energy state above the top of the upper band. These states have no analog in the 
1-chain
system or the emitter-shared 
2- and 3-chain
systems and can, since they reside outside the nearly continuous energy bands, be interpreted as bound states. This interpretation is consistent with the observation that the corresponding eigen states, which are shown in the upper left and upper right insets of Fig.~\ref{fig_energy_2ring}, are  localized in the vicinity of the cavity that is shared by the two SSH chains. The eigen state that sits below the bottom of the lower band is nodeless (upper left  inset of Fig.~\ref{fig_energy_2ring})
while the eigen state that sits above the top of the upper band is highly oscillatory, i.e., the eigen state's expansion coefficients corresponding to neighboring cavities have opposite signs
(upper right  inset of Fig.~\ref{fig_energy_2ring}).
These localized bound states are non-topological
 in nature since they have non-vanishing population in both sub-lattices and also exist for $u/v=1$
 (for this ratio, the energy gap of $\hat{H}_{\text{SSH}}$ closes) as well as periodic BCs.
 The binding energy, measured from the bottom/top of the energy band, increases---for fixed $N$---with increasing number of 
 chains.

As already alluded to, for finite 
$g/u$ and  $|\hbar \omega_e/u|\ll 1$,
the emitter state $|\text{vac};e \rangle$ has appreciable overlap
with only three of the eigen states that are located in the energy gap of the cavity-shared
2- and 3-chain
systems. 
Dashed and dotted lines in Fig.~\ref{fig_energy_ring} show the energy of the three states in the gap that have finite overlap with $|e;\text{vac} \rangle$  (top row) and the square of the overlap (bottom row) for,
respectively,
the cavity-shared 
2- and 3-chain
systems
with open BCs 
as a function of $\hbar \omega_e$ for four different $g/u$ values.
The behavior of the gap states and their energies for the 
2- and 3-chain
systems is similar to that for the 
1-chain
system, with the main feature that the avoided crossings are becoming somewhat narrower as the number of 
chains
increases from 1 to 2 and again from 2 to 3. The observed behavior is consistent with the decrease of the effective coupling constant $G$ by  factors of $1/\sqrt{2}$ and
$1/\sqrt{3}$ for the cavity-shared 
2-chain and 3-chain
systems, respectively,
relative to the 
1-chain
system 
with effective coupling constant $G$ (see Appendix~\ref{sec_appendix4}).
Since the 3-state model applies, the hybridization discussed in the previous section carries over, with the basis states being those introduced in Appendix~\ref{sec_appendix4}. The quantity $\sum_{l=1}^3 {\cal{O}}_l$ is greater than 
$0.998$
for all cavity-shared 2- and 3-chain systems considered in Fig.~\ref{fig_energy_ring}.
This 
is similar to what was discussed above for the corresponding emitter-shared 
2- and 3-chain
systems.

\section{Response to disorder}
\label{sec_ipr}

To analyze the robustness of the topological characteristics, we introduce uniformly distributed  onsite disorder of the photonic part of the Hamiltonian.
The disorder strengths $\epsilon_n$ are chosen from the disorder strength window
$[ -\Delta,\Delta]$.
In the absence of the coupling to the emitter, the onsite disorder breaks the chiral symmetry of the 
SSH part of the Hamiltonian; hopping disorder (not considered), in contrast, preserves the chiral symmetry of the
SSH part of the Hamiltonian~\cite{inui1994}.
We diagonalize the full system Hamiltonian
for a large number of onsite disorder realizations and
analyze, for each disorder realization, the three eigen states $|\psi_{l'}^{\text{gap,disorder}} \rangle$ that have the largest overlap with the 
gap states $|\psi_l^{\text{gap}} \rangle$
for the same $g/u$, $\hbar \omega_e/u$, $v/u$, and $n^*$ in the absence of disorder (recall that the three gap states are defined as the states that have energy close to zero and  depend,
for $\hbar \omega_e \ne 0$,
on the value of $g/u$).
When the disorder strength $\Delta/u$ is small, the overlap criterion employed to identify the states  $|\psi_{l'}^{\text{gap,disorder}} \rangle$ is---since the largest overlap is pretty close to $1$---``clean". For larger $\Delta/u$, in contrast, the eigen states 
$|\psi_{l'}^{\text{gap,disorder}} \rangle$
are found to deviate notably from the disorder-free gap states $|\psi_l^{\text{gap}} \rangle$; 
despite of this, the largest overlap, while notably smaller than one, allows for an ``unambiguous" identification of the 
states $|\psi_{l'}^{\text{gap,disorder}} \rangle$. s
The 
decrease of the overlap with increasing $\Delta/u$ signals that the system characteristics are fundamentally altered when the onsite disorder strength is increased.

\begin{widetext}

\begin{figure}[t]
\includegraphics[width=0.85\textwidth]{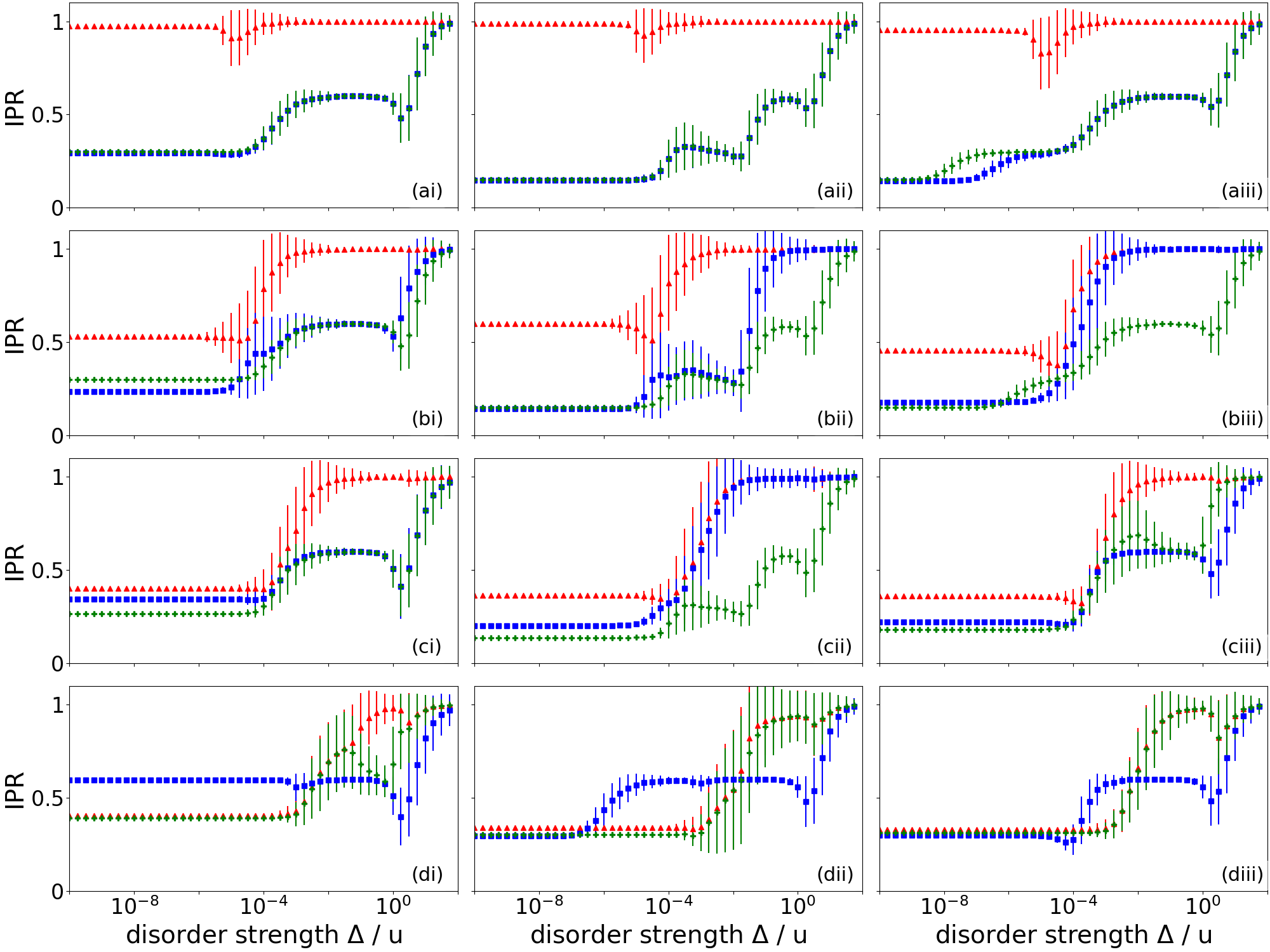}
\caption{IPR [see Eq.~(\ref{eq_IPR})] for uniformly distributed onsite disorder as a function of $\Delta/u$
(this quantity defines the scaled disorder strength window)
for   $\hbar \omega_e/u=-5\times10^{-5}$,
$N=15$, $v/u=2$, $n^*=8$, open BCs, and (ai)-(aiii) $g/u=10^{-4}$,
(bi)-(biii) $g/u=10^{-3}$,
(ci)-(ciii) $g/u=10^{-2}$, and
(di)-(diii) $g/u=10^{-1}$.
The left, middle, and right columns are for
the 
1-chain, cavity-shared 2-chain, and emitter-shared 2-chain
systems, respectively.
The IPR is calculated for the three 
eigen states of the system with disorder that have the largest overlap with the gap states
of the corresponding disorder-free system.
The symbols and error bars are obtained by averaging over $5 \times 10^3$ disorder realizations.
In panels~(ai) and (aii) as well as in parts of panels (bi) and (ci), the blue and green symbols are essentially indistinguishable.
In the large disorder regime of panels (dii) and (diii), the green and red symbols are essentially indistinguishable.
}
\label{fig_IPR}
\end{figure} 

\end{widetext}

To quantify the degree of localization of the three eigen states $|\psi_{l'}^{\text{gap,disorder}} \rangle$, 
we calculate the inverse participation ratio IPR~\cite{scollon2020},
\begin{eqnarray}
\label{eq_IPR}
\text{IPR} = \frac{\sum_{m=1}^M |c_m^{(l')}|^4 }
{\left| \sum_{m=1}^M |c_m^{(l')}|^2  \right|^2},
\end{eqnarray}
where the expansion coefficients
$c_m^{(l')}$ are given by the overlap of the state $|\psi_{l'}^{\text{gap,disorder}} \rangle$ and the $m$th site basis state, and $M$ is equal to $kN-k+2$ and $kN+1$ for the cavity- and emitter-shared cases, respectively (in this context, the state $|\text{vac};e\rangle$ is counted as one of the site basis states).
The inverse participation ratio (IPR) is a measure of localization: IPRs  
of $1$ and $0$ indicate maximal and minimal localization, respectively.
In addition, we analyze the polarization, i.e., we monitor if the states $|\psi_{l'}^{\text{gap,disorder}}\rangle$ occupy just one sub-lattice or both sub-lattices in each of the 2$k$ arms of the $k$-
chain
systems. The IPR (see Fig.~\ref{fig_IPR}) together with the polarization (not shown) quantify the robustness of the gap state characteristics against disorder.

Symbols in Fig.~\ref{fig_IPR} 
show the IPR  for the three  states $|\psi_{l'}^{\text{gap,disorder}}\rangle$ for 
chain
systems
with open BCs for
 $v/u=2$, $\hbar \omega_e/u=-5 \times 10^{-5}$, $N=15$, $n^*=8$, and
 $g/u$ values ranging from
$10^{-4}$ 
[Fig.~\ref{fig_IPR}(ai)-\ref{fig_IPR}(aiii)] to $10^{-1}$
[Fig.~\ref{fig_IPR}(di)-\ref{fig_IPR}(diii)] as a function of $\Delta/u$. The scaled disorder strength $\Delta/u$ is shown on a logarithmic scale, which
covers 12 orders of magnitude; for each disorder strength, the IPR (symbols) and error bars are calculated by averaging $5 \times 10^{3}$ independent disorder realizations.
 The colors employed in Fig.~\ref{fig_IPR} are ``matched" with those in Fig.~\ref{fig_energy_ring}, i.e., the IPRs shown in red, blue, and green coincide---in the zero disorder limit---with those for the states $|\psi_{l'}^{\text{gap,disorder}} \rangle$ whose energies and overlap squares are shown in red, blue, and green in Fig.~\ref{fig_energy_ring}. The left, middle, and right columns are for 
 the  
 1-chain
 system, the cavity-shared 
 2-chain
 system, and the emitter-shared 
 2-chain
 system, respectively. It can be seen that the changes of the IPRs and the IPRs' error bars with disorder strength depend on both the 
 chain
 geometry and the coupling strength $g/u$. 
 The IPR tends to change 
in non-trivial ways with the disorder strength,
suggesting that the disorder modifies the states
$|\psi_{l'}^{\text{gap,disorder}} \rangle$ in ways that depend intricately on the energy scales of the system.

\begin{widetext}

\begin{figure}[t]
\includegraphics[width=0.85\textwidth]{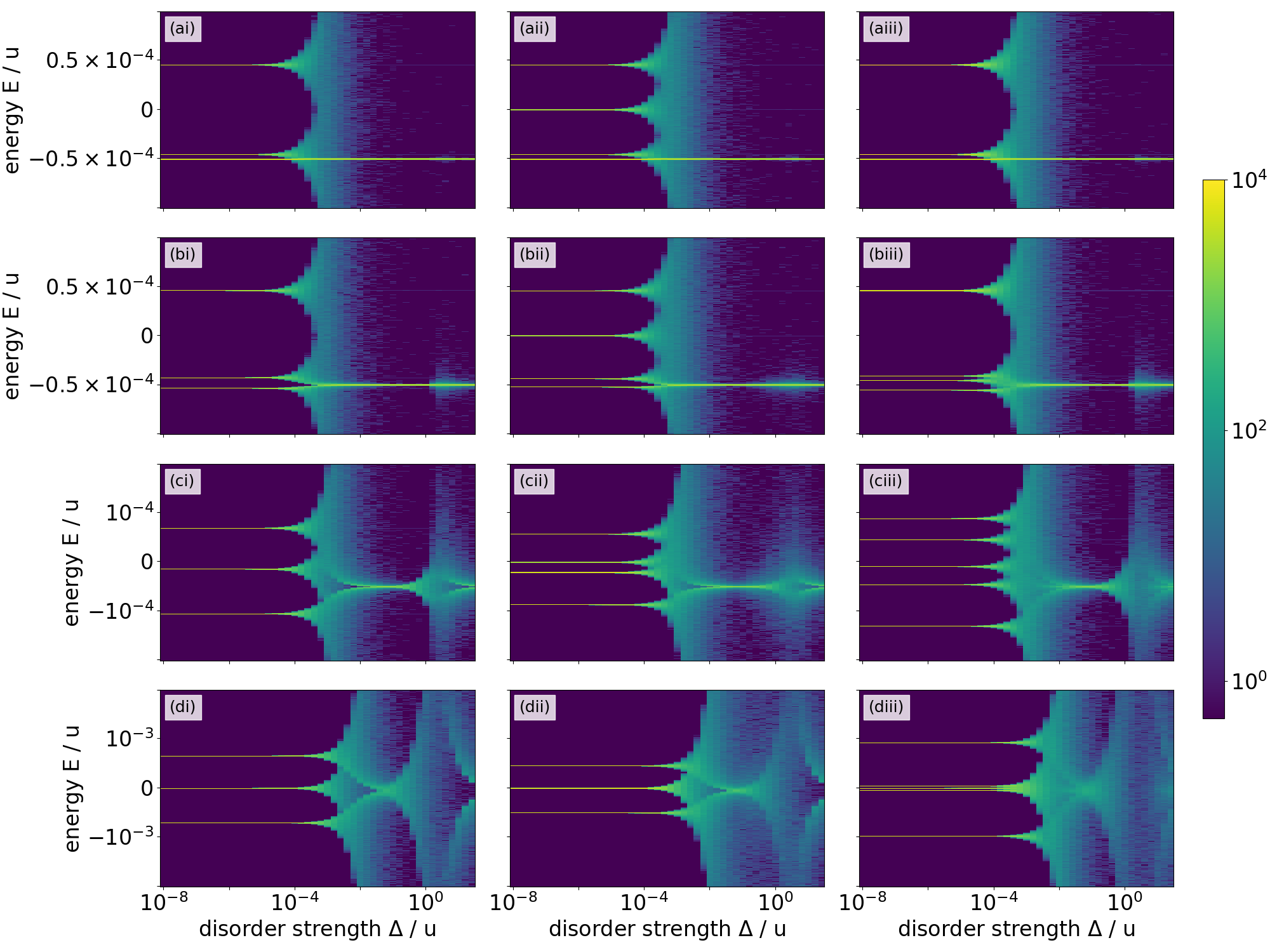}
\caption{Frequency of energy, color-coded via the scale shown on the far right,  for uniformly distributed onsite disorder 
(each panel uses $5 \times 10^3$ disorder realizations) as a function of $\Delta/u$
for   $\hbar \omega_e/u=-5\times10^{-5}$,
$N=15$, $v/u=2$, $n^*=8$, open BCs, and (ai)-(aiii) $g/u=10^{-4}$,
(bi)-(biii) $g/u=10^{-3}$,
(ci)-(ciii) $g/u=10^{-2}$, and
(di)-(diii) $g/u=10^{-1}$.
The left, middle, and right columns are for
the 
1-chain, cavity-shared 2-chain, and emitter-shared 2-chain
systems, respectively. 
}
\label{fig_energybin}
\end{figure} 

\end{widetext}

\begin{figure}[t]
\includegraphics[width=0.5\textwidth]{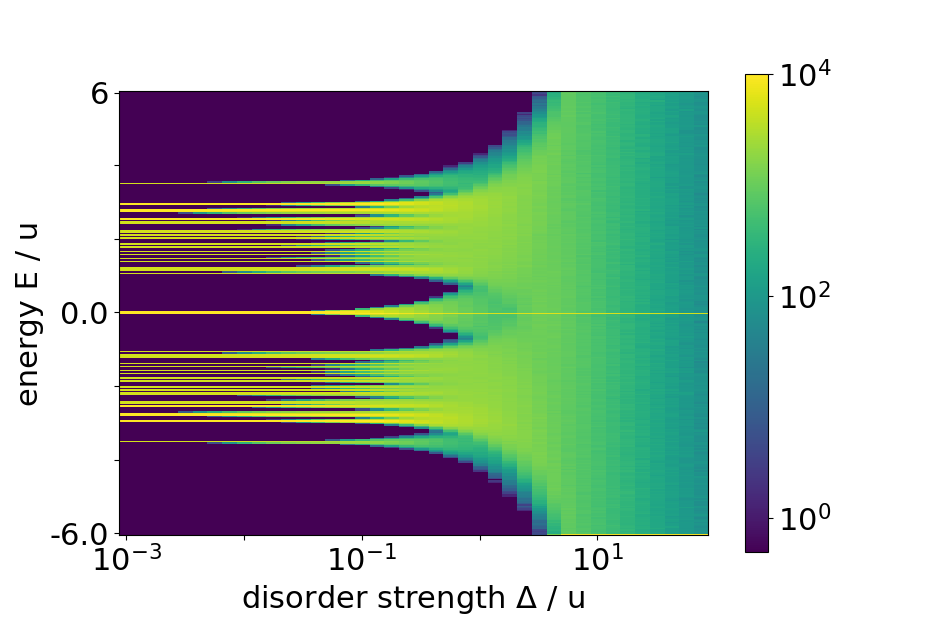}
\vspace*{0.24cm}
\caption{Same as Fig.~\protect\ref{fig_energybin}(bii), but focusing on a smaller range of disorder strengths and a larger energy window. 
}
\label{fig_energybin_blowup}
\end{figure}

Complementing the IPR, Fig.~\ref{fig_energybin} shows  the distribution of the eigen energies, averaged 
also
over $5 \times 10^3$ disorder realizations, as a function of the scaled disorder strength $\Delta/u$.
 The layout of Fig.~\ref{fig_energybin} is the same as that of Fig.~\ref{fig_IPR}, i.e., the two figures cover the same range of scaled disorder strengths, 
chain
 geometries, and coupling strengths $g/u$. For each $g/u$, the
   chosen energy window (range of the $y$-axes) in Fig.~\ref{fig_energybin} is the same as in Fig.~\ref{fig_energy_ring}. Contrary to Fig.~\ref{fig_energy_ring}, Fig.~\ref{fig_energybin} includes not only the energy of the gap states but of all eigen energies that fall into the energy window. To connect the energy distribution in the band gap with the nearly continuous energy bands, Fig.~\ref{fig_energybin_blowup} shows the distribution of eigen energies of the cavity-shared 
   2-chain
   system for a much larger energy window and somewhat smaller range of disorder strengths. The gap regime, which is the focus of Fig.~\ref{fig_energybin}, is not resolved on this scale. Plots (not shown) for the other 
   chain
   geometries and coupling strengths $g/u$ considered in this work  look essentially identical to Fig.~\ref{fig_energybin_blowup}, with the exception of the bound states below and above the energy bands, which only exist for the cavity-shared 
   $k$-chain
   systems ($k\ge 2$; see Fig.~\ref{fig_energy_2ring}). 

Combining Figs.~\ref{fig_energybin} and \ref{fig_energybin_blowup}, three regimes can be identified. 
(i) For small $\Delta/u$ ($\Delta/u \lesssim 10^{-4}$), the energies in the gap are robust to disorder, i.e., the energies in the gap are distinguishable from each other.
(ii) For intermediate $\Delta/u$ ($10^{-4} \lesssim \Delta / u \lesssim 1$), the energies that used to lie in the gap form a band that is separated from the two nearly continuous energy bands; also, the bound states are separated from the two nearly continuous energy bands. 
(iii) At large detunings ($\Delta/u \gtrsim 1$), the bands are essentially ``melted" entirely; we emphasize that there exists a state with energy  $\approx \hbar \omega_e$ for small $g/u$ and large $\Delta/u$ that is only minimally impacted by the disorder. 
Importantly, we find that the population in a given arm of the 
$k$-chain
systems is, for all  gap states $|\psi_{l'}^{\text{gap,disorder}}\rangle$, to a very good approximation either located in sub-lattice 1 or in sub-lattice 2 up to disorder strength $\Delta/u \approx 10^{-3}$
[this includes the regime (i) as well as a portion of the regime (ii) introduced above], i.e., the topological characteristic 
of population being localized on only one sub-lattice in a given arm
is preserved up to a critical disorder strength that depends relatively weakly on the coupling strength and 
chain
geometry and is about $20$ times larger than $|E_{\text{edge}}|$.

 We now discuss selected limits.
 We start with  
 the $\Delta/u \rightarrow 0$ limit (arbitrary $g/u$). 
 Appendices~\ref{sec_appendix2} and \ref{sec_appendix3} show that the
  IPRs for the 
  1-chain and emitter-shared 2-chain
  system for $\Delta/u=0$ are reproduced with high accuracy (at the percent level or better) by the analytical 3-state model expressions  [see Eqs.~(\ref{eq_ipr_1ring}) and (\ref{eq_ipr_jring_emittershared})] for all $g/u$ considered in Fig.~\ref{fig_IPR}. Notably, to approach the zero-disorder limit for the emitter-shared 
  2-chain
  system with $g/u=10^{-4}$ [Fig.~\ref{fig_IPR}(aiii)], the scaled disorder strength must be smaller than $10^{-8}$, i.e., more than three orders of magnitude smaller than $\hbar \omega_e/u$, $g/u$,
  and $|E_{\text{edge}}|/u$.
 For
  the 
  1-chain and cavity-shared 2-chain
  system, in contrast, $\Delta/u$ must be $\lesssim 10^{-5}$ for the zero-disorder limit to be approached.

  Next,
  we consider
   the small $g/u$ limit [see Figs.~\ref{fig_IPR}(ai)-\ref{fig_IPR}(aiii)
 and Figs.~\ref{fig_energybin}(ai)-\ref{fig_energybin}(aiii)].
For the smallest $g/u$ considered 
(namely, $g/u=10^{-4}$), the IPR for the
state $|\psi_{l'}^{\text{gap,disorder}} \rangle$ that is dominated by the basis state $|\text{vac};e \rangle$ is
very close to $1$ for all disorder strengths [red triangles in Figs.~\ref{fig_IPR}(ai)-\ref{fig_IPR}(aiii)]. 
The error bar is small for small $\Delta/u$, then increases, and is small again for $\Delta/u$ larger than $10^{-3}$. In the latter regime,
the state
has an energy close to $\hbar \omega_e$ [yellow-ish stripe in  Figs.~\ref{fig_energybin}(ai)-\ref{fig_energybin}(aiii)] and is localized at the emitter. Scaled disorders around $10^{-3}$ lead---for $g/u=10^{-4}$---to state localization and the reopening of an energy gap.
At very strong disorder, the IPRs of the other two states [shown in green and blue in Figs.~\ref{fig_IPR}(ai)-\ref{fig_IPR}(aiii)] also approach $1$, signaling Anderson localization~\cite{anderson1958}; this behavior is reminiscent of what was observed in Ref.~\cite{scollon2020}. 
We note that most
of the energies of these states lie outside of the energy windows shown in Figs.~\ref{fig_energybin}(ai)-\ref{fig_energybin}(aiii).

For the emitter-shared 
2-chain
system, 
a very weak disorder leads for 
$|G/E_{\text{edge}}| \ll 1$ to a change of one of the states with edge-like character [green symbols in Fig.~\ref{fig_IPR}(aiii)]: the state changes from having population in all four arms to having population in only two arms, which may belong to the same 
chain or different chains.
For the cavity-shared 
2-chain
system,
 a weak disorder leads for 
$|G/E_{\text{edge}} |\gg 1$  to a distinct change of the IPR shown by blue symbols in Fig.~\ref{fig_IPR}(dii): the state changes from being localized in two arms to being localized in one arm.
These examples illustrate that the response of the states that live in the gap to onsite disorder depends sensitively on how the SSH chains are
connected to each other (through a cavity or through an emitter).
In particular, very weak disorder can lead to distinct differences in the system response of the cavity-shared and emitter-shared 
$k$-chain
systems in a regime where the disorder modifies the eigen energies extremely weakly.
This can be understood by realizing that  the disorder breaks the discrete rotation symmetry associated with the 
$k$-chain
systems (invariance of the Hamiltonian under exchange of any two 
chain
indices for $k \ge 2$), favoring---in some cases---states that localize on one 
chain
as opposed to populating all $k$ 
chains
equally.

\section{Time-dependent signature of the hybridization}
\label{sec_dynamics}

\begin{figure}[t]
\includegraphics[width=0.47\textwidth]{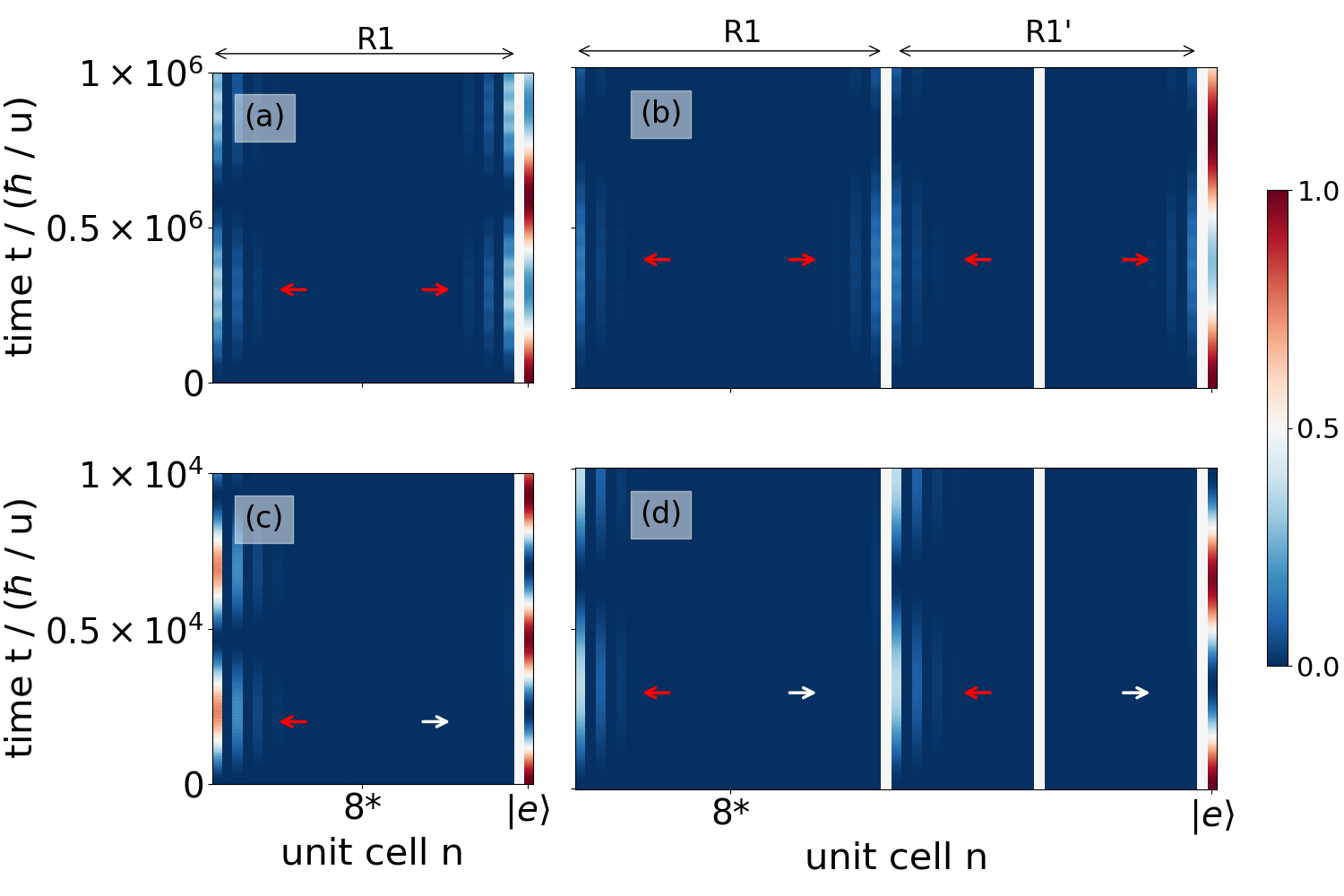}
\caption{Population dynamics of the (a)/(c) 
1-chain system and (b)/(d) cavity-shared 
2-chain
system with $N=15$, $v/u=2$, $n^*=8$, $\hbar \omega_e/u=- 5 \times 10^{-5}$, and
open BCs. The populations of the site basis
states $|n,j;\sigma\rangle$ (first 
chain of 1- and 2-chain
systems)
and
$|n',j;\sigma\rangle$ (second 
chain of 2-chain
system)
are shown as a 
function of time; the color bar shown on the right applies to all panels. The top and bottom rows 
are for $g/u=10^{-3}$ and $g/u=10^{-1}$, respectively. 
For the 
1-chain
system, the order of the basis states from left to right
is 
$|1,1;g\rangle$,
$|1,2;g\rangle$,
$\cdots$,
$|N,1;g\rangle$,
$|N,2;g\rangle$,
and
$|\text{vac};e \rangle$;
the ``white stripe" immediately to the left of the
last basis state visually separates the
basis states where the emitter is in $|g\rangle$
from those where the emitter is in
$|e\rangle$.
For the 
2-chain
system, the order of the basis states from left to right
is 
$|1,1;g\rangle$,
$\cdots$,
$|N,2;g\rangle$,
$|1',1;g\rangle$,
$\cdots$,
$|N',2;g\rangle$,
and
$|\text{vac};e \rangle$;
the first, second, and third  ``white stripes" (from left to right) visually separate the basis
states belonging to the first and second 
chain,
mark the sub-lattice~1 cavity of the second 
chain
that is ``missing" from the second 
chain
(due to it being shared with the first 
chain),
and separate basis states  where the emitter is in $|g\rangle$
from those where the emitter is in
$|e\rangle$, respectively. The red and white arrows highlight the presence and absence, respectively, of population in the cavities located at the ends of the chains.
}
\label{fig_population_dynamics}
\end{figure}

This section shows that the transition from the excited emitter state contributing predominantly to one state to contributing predominantly to two states in $k$-chain systems with open BCs can be probed dynamically (see also Ref.~\cite{bello2019}). 
Figure~\ref{fig_population_dynamics}
shows the population dynamics of the  
1-chain
system
[Figs.~\ref{fig_population_dynamics}(a) and \ref{fig_population_dynamics}(c)]
and cavity-shared 
2-chain
system
[Figs.~\ref{fig_population_dynamics}(b) and \ref{fig_population_dynamics}(d)]
as a function of the site basis 
for $\hbar \omega_e/u=-5 \times 10^{-5}$ and two different $g/u$ values, namely
$g/u=10^{-3}$ (top row) and 
$g/u=10^{-1}$ (bottom row).
The basis state
in which the emitter is excited (namely, basis state $|\text{vac};e\rangle$) is placed on the far right.
The other basis states are ordered to alternate between sub-lattice~1 and sub-lattice~2. For the 
1-chain
system, the unit cell index $n$ increases from left to right. For the 
2-chain
system, the $2N$ basis states (emitter in $|g\rangle$ and unit cell index $n$) of the first 
chain
are shown first, followed by the     
$2N-1$ basis states (emitter in $|g\rangle$ and unit cell index $n'$) of the 
second 
chain.

We first consider the dynamics of the 
1-chain
system.
The excited emitter state $|\text{vac};e\rangle$ 
has a population of $1$ at $t=0$ and then oscillates 
with a frequency that can be obtained analytically using the 3-state model introduced in 
Sec.~\ref{sec_theory_part2}.
Interestingly, while the dynamics of the 
population of state $|\text{vac};e\rangle$ is qualitatively
similar for the two $g/u$ values considered---albeit with different  oscillation frequency---,
the population dynamics of the states $|n,j;g\rangle$
shows a marked difference for the two different $g$ values.
For $g/u=10^{-3}$, both end cavities of the chain have an enhanced population when the population of the
state $|\text{vac};e\rangle$ is smallest [see red arrows in Fig.~\ref{fig_population_dynamics}(a)].
For $g/u=10^{-1}$, in contrast, the left end of the chain displays an enhanced population while the right end of the chain does not [white arrow in Fig.~\ref{fig_population_dynamics}(c)]. These population dynamics are consistent with our 
discussion in the previous section and, in particular, with our conclusion 
that the excited emitter state hybridizes with the photonic components that live on the left arm of the chain
 for $g/u \gtrsim 10^{-2}$. 
The initial state has essentially zero overlap with the 
gap state that has approximately zero energy [see blue solid  lines in Figs.~\ref{fig_energy_ssh}(d) and \ref{fig_energy_ssh}(h)] and can be, in the large $G$ limit, approximated by
$|\psi_{\text{edge,R}};g \rangle$ (see 
Sec.~\ref{sec_theory_part2} and
Appendix~\ref{sec_appendix2}). 
As a consequence, the absence of population in the right arm, marked by the white arrow in Fig.~\ref{fig_population_dynamics}(c), can be interpreted as a key fingerprint of the 
change of the hybridization of the gap states
for sufficiently large $G$. 
If we repeat the dynamical study for $\hbar \omega_e=0$ but otherwise identical parameters (not shown), we find that the photonic populations on the right arm undergo oscillations for $g/u=10^{-3}$ (the left arm has vanishing photonic populations) and those on the left arm undergo oscillations for $g/u=10^{-1}$ (the right arm has vanishing photonic populations); see the insets of Fig.~\ref{fig_new} for depictions of the corresponding $\hbar \omega_e=0$ gap states.

The dynamics of the cavity-shared 
2-chain
system is analogous to that of the 
1-chain
system, with the dynamics of the 
2-chain
system being slightly slower than that of the 
1-chain
system, as would be expected based on the small but visible changes  of the energy spectra with the  
number of chains (see the top row of
Fig.~\ref{fig_energy_ring}).
For $g/u=10^{-3}$, the population of the four end cavities  is maximal when the population of the state $|\text{vac};e\rangle$ is minimal [red arrows in Fig.~\ref{fig_population_dynamics}(b)]. 
For $g/u=10^{-1}$, in contrast, only the left ends of both chains 
get populated appreciably [red and white arrows in Fig.~\ref{fig_population_dynamics}(d)].
 As in the 
1-chain
 case, the 
 excited emitter state hybridizes with the $g=0$ edge states such that, for sufficiently strong coupling, only cavities in the left arms of the chains are occupied.
 In analogy to the 
1-chain
 case, the absence of population in two arms for sufficiently large $G$ [white arrows in Fig.~\ref{fig_population_dynamics}(d)] signals the 
 change in hybridization.
The behavior for the cavity-shared  
3-chain
system 
with open BCs
(not shown) is similar to that for the 
1-chain and cavity-shared 2-chain
systems
with open BCs.

\section{Conclusions}
\label{sec_summary}
This work investigated static and dynamic properties of one, two, and three SSH chains, 
which possess chiral symmetry, coupled weakly to a 
2LE that breaks the chiral symmetry.
In the case of a single chain, the emitter was coupled to one of the lattice sites.
In the case of the 
2- and 3-chain
systems, the emitter was either coupled to one lattice site of each 
chain
(emitter-coupled 
$k$-chain
system; $k=2$ or $3$) or to a single lattice site that 
was shared between the 
chains
(cavity-coupled 
$k$-chain
system). Since the rotating wave approximation was employed,  coupling strengths were limited to $g/u \le 10^{-1}$.
Using open BCs and working in the single-excitation manifold, this work focused on the states that reside in the energy gap between the two nearly continuous energy bands. Throughout, the  excitation energy of the emitter was chosen such that the emitter was in resonance with the band gap.
The number of states in the band was found to depend on the 
chain
geometry. For all 
chain
geometries considered, it was found that the eigen states in the gap could be grouped into three states that depend on the coupling strength $g/u$ and zero or more states that, to a very good approximation, had zero population in the emitter. A fully analytical 3-state model was  found to provide an excellent description of the $g$-dependent gap states with topological characteristics for all $k$-chain systems investigated.

This work exploited that a SSH chain 
with open BCs
is characterized by a finite edge state energy $E_{\text{edge}}$,
leading to a $g$-dependent hybridization that is absent in the system with periodic BCs. 
This finite energy scale modifies the role of the emitter from being perturbative for 
$|G/E_{\text{edge}}| \lesssim 1$
to 
being non-perturbative for 
$|G/E_{\text{edge}}| \gtrsim 1$. The hybridization of the excited emitter state and the $g=0$ edge states was analyzed in detail and the behavior was contrasted with that for the system with periodic BCs.

Generalizations and variants of the paradigmatic SSH model include studies of two nonreciprocal coupled SSH chains~\cite{cui2020}, a bipartite lattice of domain wall states~\cite{munoz2018}, and topological synchronization~\cite{wachtler2022}.
Our work adds to the growing body of emitters coupled to topological wave guides~\cite{bello2019,kim2021,barik2018,barik2020}. Extensions of the present work to 3-level emitters, which themselves support dark states, will open the door for  coupling the dark state of the emitter and the dark states of cavity-coupled 
$k$-chain
systems. 
In hyperbolic lattices, which contain 
 chain- or 
 ring-like building blocks,
 the relaxation dynamics of a 
 2LE 
 was recently proposed as a probe of the hyperbolic bath~\cite{bienias2022}.
 Another intriguing 
prospect is to work in the weak-coupling regime where the emitter plays a perturbative role and to devise sensing protocols by which the emitter dynamics, or that of two entangled emitters, can be used to probe topological matter.

{\em{Acknowledgement:}}
Support by the National Science Foundation through grant numbers  PHY-2110158 and PHY-1950235 (REU/RET) is gratefully acknowledged.

\appendix

\section{Dark states}
\label{sec_appendix1}

This appendix  determines the number of dark states, i.e., the number of eigen states of 
$\hat{H}_{\text{C}k\text{-2LE}}$ 
that have vanishing energy.
The cavity-shared 
$k$-chain
system
in the single-excitation sub-space is spanned by a total of
$(kN-k+1)+(kN)+1=2kN-k+2$ basis states.
It proves useful to reorder the site basis states as follows:
the
basis states
$1,\cdots, kN-k+1$ are of type
$|n,1;g\rangle$;
the basis states $kN-k+2,\cdots,2kN-k+1$
are of type
$|n,2;g\rangle$
(there are $kN$ basis states of this type);
and
the
basis state
$2kN-k+2$ is equal to
$|\text{vac};e\rangle$.
With this ordering,
the Hamiltonian matrix
$\underline{H}_{\text{C}k\text{-2LE}}$ 
has the simple block structure
\begin{eqnarray}
\label{eq_matrix_block}
\underline{H}_{\text{C}k\text{-2LE}}
=
\left(
\begin{array}{cc}
\underline{O} & 
\underline{V} \\
\underline{V}^{\dagger} & \underline{P}
\end{array}
\right),
\end{eqnarray}
where $\underline{O}$, $\underline{P}$, and $\underline{V}$ are matrices 
of size $(kN-k+1) \times (kN-k+1)$, $(kN+1) \times (kN+1)$,
and
$(kN+1)\times (kN-k+1)$, respectively.
The matrix elements of $\underline{O}$ are all equal to zero since  
$\hat{H}_{\text{C}k\text{-2LE}}$ 
does not couple basis states $|n,1;g\rangle$ and $|n',1;g\rangle$
for either $n=n'$ or $n \ne n'$.
The square matrix $\underline{P}$ arises from combining the states 
$|n,2;g\rangle$ and $|\text{vac};e\rangle$.
The matrix elements of $\underline{P}$ are all equal to zero, except for the element
$P_{2kN-k+2,2kN-k+2}$, which is equal to $\hbar \omega_e$. 
A key point is that the basis states are ordered such that the matrix $\underline{V}$ accounts for all ``couplings", i.e.,
the Hamiltonian terms proportional to $u$, $v$, $v_N$, and $g$ are included in $\underline{V}$.
Note in particular that, since the emitter is not coupled to sub-lattice~2 but to sub-lattice~1, the coupling constant $g$ does not enter into $\underline{P}$ (i.e., $\underline{P}$ contains at most one non-zero entry).

To proceed, we first consider the special case where the detuning $\hbar \omega_e$ is equal to zero. In this case,
all matrix elements of $\underline{P}$ are equal to zero. 
Applying the results from the Appendix of Ref.~\cite{PRB100-045414}, it follows that
the number of dark states is given by the difference, in magnitude, between the number of rows and the number of columns of $\underline{V}$.
The cavity Hamiltonian 
$\hat{H}_{\text{C}k\text{-2LE}}$ 
hence supports $k$ dark states for $\hbar \omega_e=0$.
For $\hbar \omega_e \ne 0$, one of the dark states turns ``bright", i.e., the energy 
of this state is pushed away from zero; this behavior is clearly visible in the top row of Fig.~\ref{fig_energy_ring}.
Mathematically, the disappearance of the dark state follows since the right lower matrix element of $\underline{P}$ takes on a finite value. 
To summarize,  the cavity-coupled 
$k$-chain
systems 
with $\hbar \omega_e \ne 0$ support  $k-1$ dark states (see Fig.~\ref{fig_schematic_ring}).

The arguments for determining the number of dark states for the emitter-shared 
$k$-chain
systems proceed analogously.
Keeping the same grouping of the basis states,
the matrices
$\underline{O}$, $\underline{P}$, and $\underline{V}$ are of size $(kN)\times(kN)$, $(kN+1) \times (kN+1)$,
and
$(kN+1)\times (kN)$, respectively.
The key difference compared to the cavity-shared systems is that the number of basis states of type $|n,1;g\rangle$ is $kN$ in the emitter-shared system as opposed to $kN-k+1$.
Applying the results  
from the Appendix of Ref.~\cite{PRB100-045414}, it follows that the emitter-shared 
$k$-chain
systems support exactly one dark state if and only if $\hbar \omega_e=0$. 
Figure~\ref{fig_schematic_ring} indicates the absence of dark states since it summarizes the more
general
$\hbar \omega_e \ne 0$ case (the special $\hbar \omega_e=0$ case is referred to in the caption).

\section{3-state model for gap states of 
1-chain
system}
 \label{sec_appendix2}

 Section~\ref{sec_theory_part2}  
 discusses
 selected properties of the 3-state Hamiltonian 
$\underline{H}_{\text{3-st.}}(G)$.
 This appendix presents analytical expressions for three special cases.
 The eigen values of $\underline{H}_{\text{3-st.}}(G)$ can be obtained by solving the cubic
 equation
 \begin{eqnarray}
 \lambda^3 - \hbar \omega_e \lambda^2 - \left[ 2 G^2 + 
 (E_{\text{edge}})^2 \right] \lambda \nonumber + \\  
 \hbar \omega_e (E_{\text{edge}})^2=0.
 \end{eqnarray}

 Special case~1: When the detuning vanishes,
 the eigen energies are equal to 
 \begin{eqnarray}
 \lambda|_{\hbar \omega_e=0} =0
 \end{eqnarray} 
 and 
 \begin{eqnarray}
 \lambda|_{\hbar \omega_e=0}=
\pm \left[ 2 G^2 + 
 (E_{\text{edge}})^2 \right]^{1/2}.
\end{eqnarray} 
This shows that the splitting between the energetically lowest- and highest-lying gap states is approximately equal to
$2|E_{\text{edge}}|$ and $2 \sqrt{2}  G$ when 
 $2 G^2 \ll 
 (E_{\text{edge}})^2$
 and 
 $2 G^2 \gg 
 (E_{\text{edge}})^2$, respectively.
 These inequalities suggest that a ``transition" occurs when
 $\sqrt{2} G$ is comparable to $|E_{\text{edge}}|$.
The unnormalized zero-energy eigen state reads $(G/E_{\text{edge}},-G/E_{\text{edge}},1)$.
 
 Special case~2:
Figure~\ref{fig_energy_ring} shows that the energy levels undergo two separate avoided crossings
when $g/u$ is small [Figs.~\ref{fig_energy_ring}(a) and \ref{fig_energy_ring}(b)]. When $g/u$ is comparatively large [Fig.~\ref{fig_energy_ring}(d)],
in contrast, the two avoided crossings can no longer
be treated separately.
To identify the energy scale at which the crossings start to overlap, we
consider
 a  2-state model, which removes the first row and first column from
$\underline{H}_{\text{3-st.}}(G)$.
 The eigen energies $\lambda_{\text{2-st.}}$ of the 2-state
 model are given by
 \begin{eqnarray}
 \lambda_{\text{2-st.}} = 
 \frac{\hbar \omega_e + E_{\text{edge}}}{2}
 \pm 
 \sqrt{
 \left(
 \frac{\hbar \omega_e - E_{\text{edge}}}{2}
 \right)^2
 + G^2 }
 \end{eqnarray}
 and the energy splitting at the avoided crossing is equal to
 $2 G$. Since the energies of 
 these two states have the same magnitude but opposite sign, 
 it is readily argued from this splitting that the two avoided crossings can no longer be treated separately if 
 $2  G$ approaches $|E_{\text{edge}}|$. As expected, the 
 ``competition scale" 
 obtained via the 2-state model is similar to that obtained from the 3-state model.
 For the data shown in Fig.~\ref{fig_energy_ring}, the 
 competition
 scale is reached roughly when $g/u$ is equal to $10^{-2}$, consistent with 
 what is concluded by visual inspection.
 
 Special case~3:
 To gain additional insight into the larger $g/u$ regime [Fig.~\ref{fig_energy_ring}(d)], 
 we return to the 3-state model and consider the limit where $|E_{\text{edge}}|$ is much smaller than 
 $G$. Setting $E_{\text{edge}}=0$,
 the eigen values of $\underline{H}_{\text{3-st.}}(G)$ are 
 \begin{eqnarray}
 \label{eq_lambda0_edge0}
 \lambda |_{E_{\text{edge}}=0}=0
 \end{eqnarray}
 and
 \begin{eqnarray}
 \lambda |_{E_{\text{edge}}=0}=
 \frac{\hbar \omega_e}{2} \pm \sqrt{
 \left(
 \frac{\hbar \omega_e}{2}
 \right)^2
 + 2 G^2
 }.
 \end{eqnarray}
 The eigen state corresponding to $\lambda |_{E_{\text{edge}}=0}=0$
 is 
 equal to
 $(|\psi_{+}^{\text{C}1}\rangle - |\psi_{-}^{\text{C}1}\rangle )|g\rangle/\sqrt{2}=|\psi_{\text{edge,R}}\rangle|g\rangle$, i.e., this eigen state has non-vanishing amplitude only in one side of the 
 chain
 and only in sub-lattice~2
 (see Sec.~\ref{sec_theory_part2} for further discussion).
  
 Altogether, the analysis outlined in this appendix
 shows that the emitter acts as a perturbation
 when $g/u$ is much smaller than about $|E_{\text{edge}}/(u  c_{n^*,1})|$. 
 The state $|\text{vac};e\rangle$ hydridizes with the $g=0$ edge states (emitter in $|g\rangle$) when the effective coupling $G$ is comparable to $|E_{\text{edge}}|$. Where the transition occurs
 can be tuned 
 by increasing $g/u$ for fixed $N$ or by increasing $N$ for fixed $g/u$. 
 We note that the unit cell $n^*$ at which the emitter is placed can also be used as a tuning knob.

 We now present approximate analytical expressions for the IPR. We start with the photonic Hamiltonian (excluding the emitter Hilbert space).  
 The IPR for the 
 states $|\psi_{\pm}^{\text{C}1}\rangle$ 
 reads
 \begin{eqnarray}
  \label{eq_ipr_edge}
  \text{IPR}_{|\psi_{\pm}^{\text{C}1}\rangle}=
  \frac{{\cal{N}}^2 \left[ 1+\left(\frac{u}{v}\right)^{2N} \right]}{2 +2\left(
  \frac{u}{v}\right)^2},
 \end{eqnarray} 
which can be simplified to
  \begin{eqnarray}
  \label{eq_ipr_approx}
  \text{IPR}_{|\psi_{\pm}^{\text{C}1}\rangle}
  \approx \frac{1-\left( \frac{u}{v} \right)^2}{2+2\left( \frac{u}{v}\right)^2}.
  \end{eqnarray}
For $u/v=2$ and $N=15$, e.g., Eq.~(\ref{eq_ipr_approx})
evaluates to
$3/10$, which deviates from Eq.~(\ref{eq_ipr_edge}) by less than $6 \times 10^{-10}$.

To obtain approximate analytical expressions for the IPR for the states
$|\psi_l^{\text{gap}} \rangle$ 
(1-chain
system with finite $g$), we write
\begin{eqnarray}
\label{eq_gap_state_expand}
 |\psi_l^{\text{gap}} \rangle= d_+^{(l)} | \psi_+^{\text{C}1}\rangle |g\rangle+
 d_-^{(l)} | \psi_-^{\text{C}1}\rangle |g\rangle
 + \nonumber \\
 d_e^{(l)} |  \text{vac},e\rangle,
\end{eqnarray}
where the expansion coefficients
$d_+^{(l)}$, $d_-^{(l)}$,
and $d_e^{(l)}$
are extracted from the eigen vectors of the 3-state model.
Evaluating the IPR for the states given in Eq.~(\ref{eq_gap_state_expand}), we find
\begin{eqnarray}
\label{eq_ipr_1ring}
 \text{IPR}_{|\psi_l^{\text{gap}}\rangle}
 = \nonumber \\
 \text{IPR}_{|\psi_{\pm}^{\text{C}1}\rangle}
 \left(
 |d_+^{(l)}|^4 +
 |d_-^{(l)}|^4
 +
6 |d_+^{(l)}|^2 |d_-^{(l)}|^2
 \right)+ 
  |d_e^{(l)}|^4.
\end{eqnarray} 

\section{Emitter-shared 2- and 
3-chain
systems}
\label{sec_appendix3}

To construct a few-state model that describes the $g$-dependence of the energy levels that lie in the middle of the gap for $k>1$, we first consider the emitter-shared 
2-chain
system. We introduce approximate expressions for the $g=0$ eigen states with energy close to zero.
Since the two 
chains
are decoupled for $g=0$, the system supports two eigen states with energy $E_{\text{edge}}$
and two eigen states with energy $-E_{\text{edge}}$,
\begin{eqnarray}
 |\psi_{\pm}^{\text{C}1} \rangle | \text{vac,C}2 \rangle |g \rangle
\end{eqnarray}
and 
\begin{eqnarray}
 | \text{vac,C}1 \rangle |\psi_{\pm}^{\text{C}2} \rangle |g \rangle,
\end{eqnarray}
where $|\text{vac,C}k \rangle$
refers to the vacuum state of 
chain
$k$ and where the state $|\psi_{\pm}^{\text{C}2} \rangle$ is defined analogously to $|\psi_{\pm}^{\text{C}1} \rangle$ (see 
Sec.~\ref{sec_theory_part2}).
To construct a few-state model, we form linear combinations of the two states that have energy $E_{\text{edge}}$ as well as   linear combinations of the two states that have energy $-E_{\text{edge}}$:
\begin{eqnarray}
 \frac{1}{\sqrt{2}}
 \left(
 |\psi_{+}^{\text{C}1} \rangle | \text{vac,C}2 \rangle
 \pm
 | \text{vac,C}1 \rangle |\psi_{+}^{\text{C}2} \rangle
 \right) |g \rangle
\end{eqnarray}
and
\begin{eqnarray}
 \frac{1}{\sqrt{2}}
 \left(
 |\psi_{-}^{\text{C}1} \rangle | \text{vac,C}2 \rangle
 \pm
 | \text{vac,C}1 \rangle |\psi_{-}^{\text{C}2} \rangle
 \right) |g \rangle.
\end{eqnarray}
The ``$+$"-linear combinations couple to the state $|\text{vac,C}1\rangle | \text{vac,C}2\rangle|e\rangle$
while the ``$-$"-linear combinations do not.
Correspondingly, we consider a 3-state model that is spanned by the two ``$+$"-linear combinations and $|\text{vac,C}1\rangle | \text{vac,C}2\rangle|e\rangle$.
Calculating the coupling matrix elements, we find that the coupling strength is $\sqrt{2}$-times larger than that of the 
1-chain
system, i.e., the 3-state Hamiltonian is given by
$\underline{H}_{\text{3-st.}}(\sqrt{2}G)$ [Eq.~(\ref{eq_1ring_matrix}) with $G$ replaced by $\sqrt{2}G$].

The emitter-shared 
3-chain
system supports seven states with energy close to zero. Forming appropriate linear combinations, we find that only three of these are shifted when $g$ is turned on. Thus, we construct a 3-state model spanned by the states
 \begin{eqnarray}
 \frac{1}{\sqrt{3}}
 {(} 
  |\psi_{+}^{\text{C}1} \rangle
  | \text{vac,C}2\rangle
  | \text{vac,C}3\rangle
  + \nonumber \\
 | \text{vac,C}1\rangle|\psi_{+}^{\text{C}2} \rangle
 | \text{vac,C}3\rangle
 +\nonumber \\
 | \text{vac,C}1\rangle| \text{vac,C}2\rangle
 |\psi_{+}^{\text{C}3} \rangle
  {)}
  |g\rangle,
 \end{eqnarray}
 \begin{eqnarray}
 \frac{1}{\sqrt{3}}
 ( 
  |\psi_{-}^{\text{C}1} \rangle
  | \text{vac,C}2\rangle
  | \text{vac,C}3\rangle
  + \nonumber \\
 | \text{vac,C}1 \rangle|\psi_{-}^{\text{C}2} \rangle
 | \text{vac,C}3\rangle
 +\nonumber \\
 | \text{vac,C}1\rangle| \text{vac,C}2\rangle
 |\psi_{-}^{\text{C}3} \rangle
  )
  |g\rangle,
 \end{eqnarray}
 and
 \begin{eqnarray}
  | \text{vac,C}1\rangle| \text{vac,C}2\rangle| \text{vac,C}3\rangle|e\rangle.
 \end{eqnarray}
 The
 3-state Hamiltonian for the emitter-shared 
 3-chain
 systems is given by
$\underline{H}_{\text{3-st.}}(\sqrt{3}G)$ [Eq.~(\ref{eq_1ring_matrix}) with $G$ replaced by $\sqrt{3}G$]. For the emitter-shared 
$k$-chain
system, the effective coupling energy is $\sqrt{k}G$.
 
 As in the 
 1-chain system, we can---analogously to Eq.~(\ref{eq_gap_state_expand})---write the gap states $|\psi_l^{\text{gap}} \rangle$ for the emitter-shared 
$k$-chain
 systems as a superposition of the three states that span the 3-state Hamiltonian.
Evaluating the IPR within the 3-state model,  we find
\begin{eqnarray}
\label{eq_ipr_jring_emittershared}
 \text{IPR}_{|\psi_l^{\text{gap}}\rangle}
 = \nonumber \\
 \frac{\text{IPR}_{|\psi_{\pm}^{\text{C}1}\rangle}}{j}
 \left(
 |d_+^{(l)}|^4 +
 |d_-^{(l)}|^4
 +
6 |d_+^{(l)}|^2 |d_-^{(l)}|^2
 \right)+ 
 |d_e^{(l)}|^4,
\end{eqnarray} 
 where $d_+^{(l)}$, $d_-^{(l)}$, and $d_e^{(l)}$ are obtained from the eigen vectors of the 3-state Hamiltonian.
 
\section{Cavity-shared 2- and 
3-chain
systems}
\label{sec_appendix4}
 
For the
cavity-shared 
$k$-chain
systems, the $g=0$ eigen states with eigen energy close to zero 
fall into three groups.
The first group contains, for the 
2-chain system (3-chain
system), two (four) states with
energies that are finite but different from $\pm E_{\text{edge}}$
and
that are, to a very good approximation, not affected when the coupling $g$ is turned on. 
The second group contains one (two) non-topological dark states (see Appendix~\ref{sec_appendix1}).
The third group contains three states 
with energies $-E_{\text{edge}}$,
$E_{\text{edge}}$, and $\hbar \omega_e$ that couple to the emitter when $g$ is non-zero.
The states with energies $\pm E_{\text{edge}}$ are essentially identical to those introduced in Appendix~\ref{sec_appendix3}, with the exception that there only exists one basis state
$|n^*,1\rangle$
as opposed to $k$ basis states
$|n^*,1; \text{C}k\rangle$.
Using these two states together with $|\text{vac};e\rangle$, we find that
 the 
 3-state
 Hamiltonian
matrix 
for the cavity-shared systems is identical to
that for the 1-chain systems
but with reduced coupling constant [Eq.~(\ref{eq_1ring_matrix}) with $G$ replaced by $G/\sqrt{k}$].
The reduction of the coupling energy compared to the 
1-chain
and emitter-shared systems is due to the fact that the 
cavity that the emitter is coupled to is shared among all 
chains. 
\end{document}